%% file: main.tex
\documentclass[lettersize,journal]{IEEEtran}
\usepackage[noadjust]{cite}
\usepackage{textcomp}
\def\BibTeX{{\rm B\kern-.05em{\sc i\kern-.025em b}\kern-.08em
    T\kern-.1667em\lower.7ex\hbox{E}\kern-.125emX}}

\usepackage{algorithm}
\usepackage{algpseudocode}
\usepackage{amsmath,amsthm}
\usepackage{amssymb}
\usepackage{graphicx}
\usepackage{subcaption}
\usepackage[usenames,dvipsnames,table]{xcolor}
\usepackage{braket}
\newtheorem{theorem}{Theorem}[section]
\newtheorem{remark}{Remark}[theorem]

\newcommand{\prs}{PRIMARY-RATESUM}
\newcommand{\prf}{PRIMARY-RATEFAIR}
\newcommand{\rrs}{REFLECTION-RATESUM}
\newcommand{\rrf}{REFLECTION-RATEFAIR}

\title{Scheduling in Quantum Satellite Networks: Fairness and Performance Optimization}

\author{Ashutosh Jayant Dikshit, Naga Lakshmi Anipeddi, Prajit Dhara, Saikat Guha, Deirdre Kilbane, Leandros Tassiulas, Don Towsley, and Nitish K. Panigrahy*
\thanks{*Corresponding author (npanigrahy@binghamton.edu).}
\thanks{A. Dikshit and N. Panigrahy  are with Binghamton University, USA.}
\thanks{P. Dhara was with University of Arizona, USA at the time of this work.}
\thanks{N. Anipeddi and D. Kilbane are with South East Technological University, Ireland.}
\thanks{S. Guha is with University of Maryland, USA.}
\thanks{L. Tassiulas is with Yale University, USA.}
\thanks{D. Towsley is with University of Massachusetts Amherst, USA.}
}

\begin{document}

\maketitle
\input{abstract}
\input{intro.tex}
\input{prelims.tex}

\input{optimization.tex}
\input{reflector.tex}
\input{simulations.tex}
\input{conclusion.tex}
\bibliographystyle{IEEEtran}
\bibliography{bib}

\end{document}

%% file: abstract.tex
\begin{abstract}
Quantum satellite networks offer a promising solution for achieving long-distance quantum communication by enabling entanglement distribution across global scales. This work formulates and solves the quantum satellite network scheduling problem by optimizing satellite-to-ground station pair assignments under realistic system and environmental constraints. Our framework accounts for limited satellite and ground station resources, fairness, entanglement fidelity thresholds, and real-world non-idealities including atmospheric losses, weather and background noise. In addition, we incorporate the complexities of multi-satellite relays enabled via inter-satellite links. We propose an integer linear programming (ILP) based optimization framework that supports multiple scheduling objectives, allowing us to analyze tradeoffs between maximizing total entanglement distribution rate and ensuring fairness across ground station pairs. Our framework can also be used as a benchmark tool to measure the performance of other potential transmission scheduling policies.
\end{abstract}

%% file: intro.tex
\section{Introduction}
Quantum entanglement distribution over long distances is necessary for numerous quantum applications such as quantum key distribution (QKD) \cite{ekert1991quantum}, quantum sensing \cite{gottesman2012longer, komar2014quantum, eldredge2018optimal}, teleportation \cite{Bennett93}, and distributed quantum computing \cite{cirac1999distributed}. Two promising solutions for global-scale entanglement distribution are ground-based fiber links with quantum repeaters, and satellite-based free space links.

In a ground-based quantum network, entangled photonic qubits are transmitted through optical fiber links connecting remote users. Long distance entanglement is achieved by inserting a chain of quantum repeaters \cite{Dur99} in between the remote users. An alternative strategy is to use satellite-based free space links \cite{Gundogan21, Chiti24, deParny23, Wallnofer22}. In optical satellite links, the majority of photon's propagation path is in empty space, thus incurring much less channel loss and decoherence as compared to a ground based fiber link.  Additionally, recent experimental successes in satellite based quantum communication \cite{Liao17, Ren17}, have made it possible to envision a global scale satellite-based quantum network.

In satellite-based quantum network, the goal is to generate entanglements between multiple pairs of ground stations. Each satellite is equipped with photon sources that generate and distribute pairs of entangled photons to ground stations. While entanglement distribution between two ground stations using a single satellite is important, achieving global coverage and supporting multiple ground station pairs require deploying a satellite constellation \cite{Khatri21}. 

Effectively managing transmissions in such a constellation introduces a key challenge: determining which satellite should serve which ground station pairs and at what time, often referred to as the {\it scheduling problem} in quantum satellite networks \cite{panigrahy2022optimal, williams2024scalable,maule2024fair, wang2023exploiting, chang2023entanglement, wei2024optimizing}. While generating a schedule, one also needs to keep in mind that both satellites and ground stations have limited resources. An optimal schedule assigns satellites to ground station pairs and coordinating their timing, so that resources are not overbooked and the network can maximize a certain objective, for example generate as many entanglements as possible or prioritize fair distribution of entanglements.

Unlike classical satellite scheduling, where visibility is often considered for individual ground stations, quantum entanglement distribution typically requires simultaneous line of sight visibility between the satellite and both ground stations, along with meeting a desired entanglement quality\footnote{Quality of a distributed entanglement is often measured using fidelity.} threshold. Additionally, factors such as weather, cloud cover, atmospheric conditions, and  background photon noise during daytime can significantly degrade the delivery rate and quality of entanglements. An effective scheduling strategy must take into account these non-idealities when optimizing overall network objectives.

Recent work has explored the feasibility of relaxing the simultaneous line of sight visibility requirement. When a ground station pair is not simultaneously visible to a single satellite, entanglement distribution can be achieved via additional satellites using inter-satellite links \cite{goswami2023satellite, huang2024vacuum, gu2025quesat}. While this multi-satellite relay improves network connectivity, it also increases the solution space, making the scheduling problem combinatorially more complex.

In this work, we formulate and solve the scheduling problem in quantum satellite networks by determining optimal satellite to ground station pairings while accounting for limited resources, fairness, entanglement quality thresholds, real world non-idealities such as atmospheric and background noise, and the added complexity introduced by multi-satellite relays that use inter-satellite links. To the best of our knowledge, existing studies have addressed only one or a subset of the above issues. Our contributions are summarized below.

\begin{itemize}
    \item We develop an integer linear programming based optimization framework for solving the quantum satellite network scheduling problem. We optimize different scheduling objectives and show the tradeoff between maximizing aggregate entanglement distribution rate and fairness among different ground station pairs.
    
    \item We model realistic entanglement generation sources based on Spontaneous Parametric Down-Conversion (SPDC) and account for imperfections both at the source and at the receivers.
    
    \item We comprehensively model the non-idealities introduced by atmospheric conditions, such as weather variations, cloud coverage, and the presence of background photons, showing their significance in affecting fidelity and distribution rates of entanglements.
    
    \item Finally, we develop an optimal scheduling strategy for a two-satellite relay scenario, where two satellites jointly coordinate to facilitate entanglement distribution between a pair of ground stations.
    
\end{itemize}

The rest of the paper is organized as follows. In Section \ref{sec:model}, we  introduce the system model relevant to this work. Section \ref{sec:assignment} discusses different optimization objectives associated with the quantum satellite scheduling problem. Section \ref{sec:reflection} presents discussions related to the two-satellite relay based formulation associated with the scheduling problem. We present simulation results in Section \ref{sec:numeric}. Section \ref{sec:conclusion} concludes the paper and discusses some possible future work. 

%% file: prelims.tex
\section{System Model}\label{sec:model}
\begin{figure*}[!htbp]
\centering
\begin{minipage}{0.65\textwidth}
\includegraphics[width=1\textwidth]{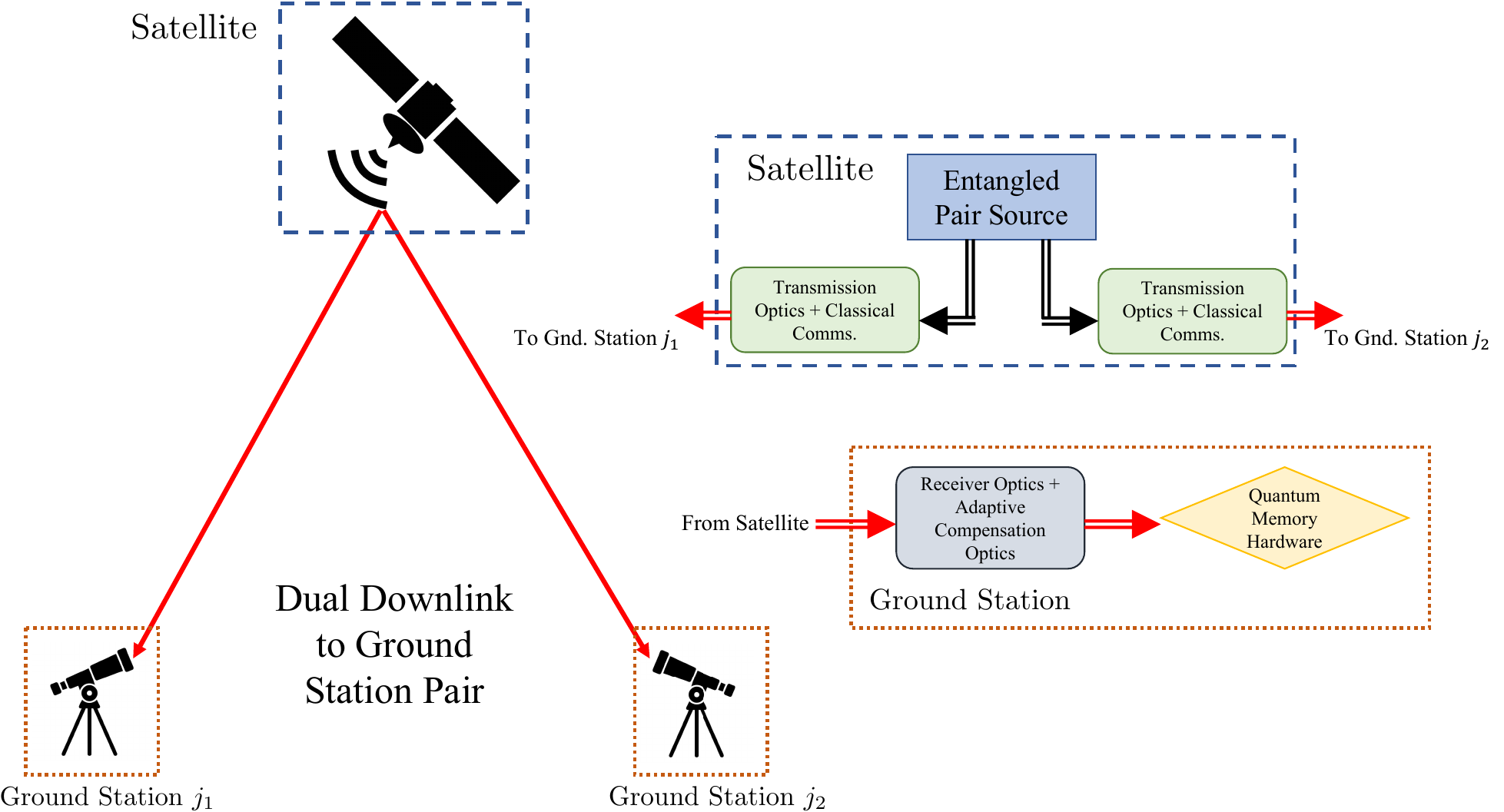}
\vspace{0.1cm}
\end{minipage}
\caption{Dual downlink architecture for photonic entanglement distribution. The satellite platform consists of spontaneous parametric down conversion (SPDC) based entangled pair sources that produce the state in Eq.~\eqref{eqn:srcnative} and distribute each qubit to the ground stations, with the necessary transmission optics. The ground stations consists of receiver optics and adaptive optics (to minimize atmospheric distortion) that couple into the quantum memory hardware. We assume that the system has perfect timing synchronization, accurate pointing and tracking, and inter-station terrestrial classical communication links that are required for the protocol to succeed.}
\label{dualdownlink}
\end{figure*}
In this section, we introduce the system model used in the rest of the paper.
\subsection{Satellite constellations and ground station pairs}\label{subsec:constellation}
We consider a satellite based quantum network that distributes shared entangled photonic quantum states between a set of ground station pairs. Denote by $S$, $G$, and $F$ the sets of satellites, ground stations, and pairs of ground stations that request entanglement respectively. The ground stations are located on earth, which rotates around its axis with a rotation period of $24$ hours. The satellites are deployed in constellations orbiting around the earth at certain altitudes. Several examples of constellations that have been considered for classical and quantum communications include Polar, Walker, Iridium, Starlink, and Kuiper \cite{Khatri21}. While our results are applicable to any constellation, for simulation purposes, we will use the polar constellation. A complete simulation study and comprehensive comparison of other constellations remains part of our future work.

We assume time is divided into fixed length time slots. Both satellites and ground stations have limited numbers of resources, such as transmitters, photon sources and receivers. Thus at the beginning of each time slot, based on current positions of the satellites and the ground stations, the satellites  are allocated to ground station pairs to maximize the overall system-wide entanglement distribution rate subject to resource constraints. We discuss this in detail in Section \ref{sec:assignment}.

\subsection{Dual Downlink Entanglement Distribution}\label{sub:dual-downlink}
In a satellite based quantum network, the satellites have entanglement sources that generate and distribute entanglements across ground stations. Each satellite has one or more photon sources that produce entangled pair of photons. Satellites send each of the photons to a pair of ground stations using a down-link channel. After an entangled pair is successfully received by each of the two ground stations, it can be used by any quantum application such as QKD or quantum teleportation.

We denote the elevation angle limit to be $\theta_e$. Thus, the entangled photon can successfully be received at the ground station as long as the elevation angle (angle between the satellite and horizon at the ground station) of a satellite from the horizon exceeds $\theta_e$. The elevation angle is used to account for terrain obstructions, such as buildings and mountains. For successful entanglement distribution by a satellite, the elevation angles for the same satellite from both ground stations must exceed $\theta_e$.

\subsection{Entangled Pair Source}
For the dual downlink configuration highlighted in Section \ref{sub:dual-downlink}, we utilize a spontaneous parametric down-conversion based dual-rail polarization entangled pair source. This source is well-studied and widely utilized for other tasks that involve quantum entanglement generation~\cite{Krovi2016-my,Kok2000-ml}. The output quantum state from such a source is described in the photon number Fock basis (up to the support of $2$ photon pairs)~\cite{Kok2000-ml}:

\begin{align}
		\ket{\psi^\pm}&\!=   N_0 \left[\!\sqrt{p(0)} \ket{0,0;0,0}\!  +\! \sqrt{\frac{p(1)}{2}} \left(\ket{1,0;0,1}\pm\ket{0,1;1,0} \right) \right.\nonumber\\
		&\left. +  \sqrt{\frac{p(2)}{3}}\left(\ket{2,0;0,2}\pm \ket{1,1;1,1}+\ket{0,2;2,0}\right)\right],
		\label{eqn:srcnative}
\end{align}
where $N_0$ is a normalization constant and the term coefficients $ p(n) $ are given by,
\begin{equation}
p(n) = (n+1)\frac{{N_s}^n}{(N_s + 1)^{n+2}}. 
\label{eqn:geometric_dist}
\end{equation}
with $N_s$ denoting the mean photon number per mode of the state. The entangled pair (in the dual rail basis) component of the above state is $\left(\ket{1,0;0,1}\pm\ket{0,1;1,0} \right) \sqrt{2}$; the vacuum portion is $\ket{0,0;0,0}$, and all other terms are spurious two-pair emission terms. By controlling the $N_s$ parameter one can limit the emission of these two-pair terms whilst making sure the output state has a considerable amount of the utilizable entanglement~\cite{Krovi2016-my}. 

\subsection{Channel Loss}
Since the link involves free-space optical transmission, any analysis of such links must take into account the characteristics of the optical channel. For the present study, we account for transmission loss by studying the property of the state after it passes through a bosonic pure loss channel. Each qubit (i.e. pair of modes) undergoes loss that is parametrized by a channel transmissivity $ \eta_{ig}(t) $ for a channel between satellite $ i $ and ground station $g$ at time $t$. Details of the $\eta_{ig}(t) $ calculation  is given as follows. 

\begin{figure}[!htbp]
\centering
\begin{minipage}{0.5\textwidth}
\includegraphics[width=1\textwidth]{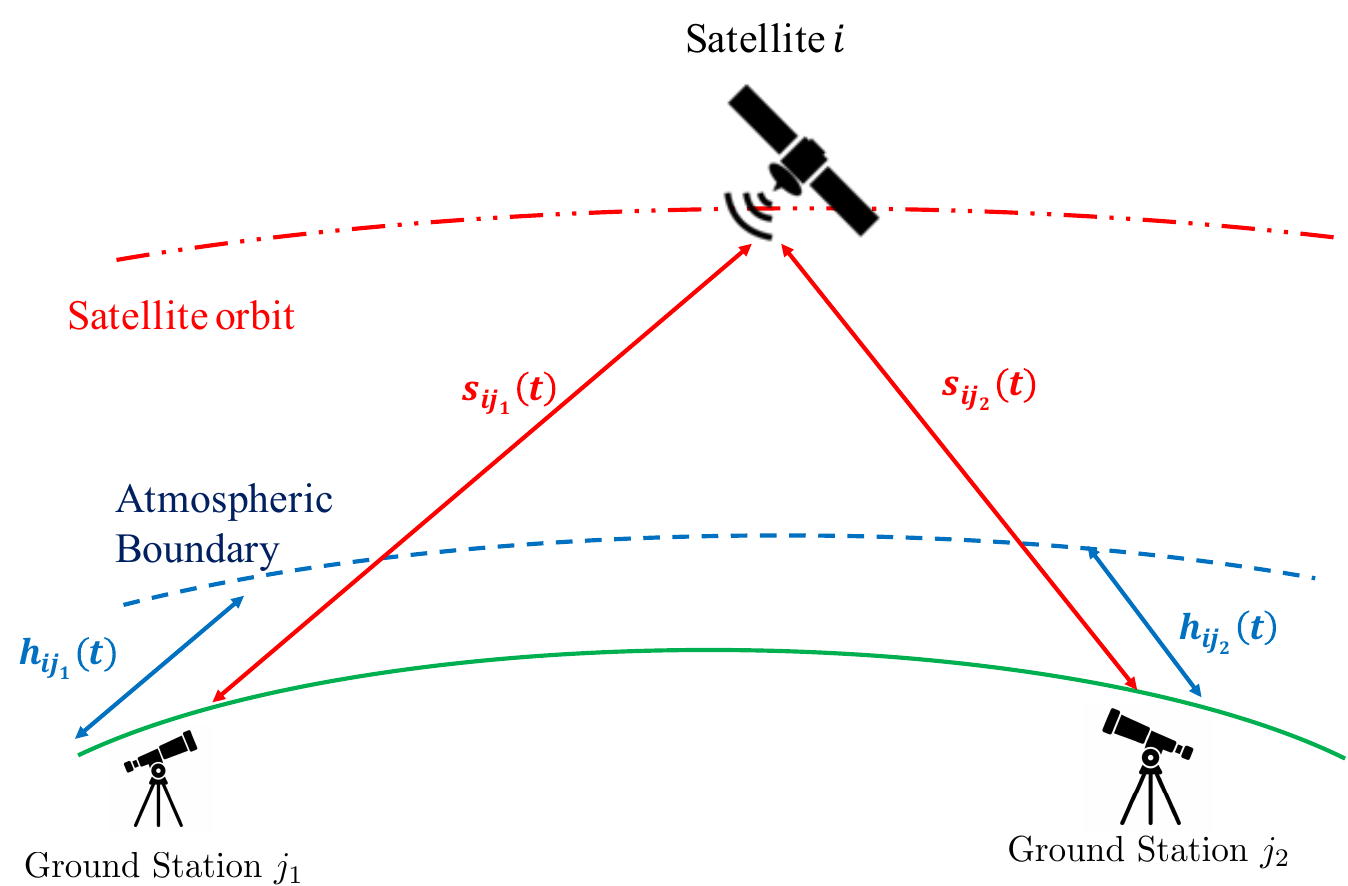}
\end{minipage}
\caption{Total (Free space + atmospheric) and atmospheric distances of Satellite-to-ground station channels at time $t$.}
\label{sat-dist}
\end{figure}

The transmissivity $ \eta_{ig}(t) $ for a channel between satellite $i$ and ground station $g$ at time $t$ depends on the free space and atmospheric distances as shown in Figure \ref{sat-dist}. Let $s_{ig}(t)$ be the distance between satellite $i$ and ground station $g$ at time $t$. Denote by $h_{ig}(t)$ the distance between ground station $g$ and atmospheric boundary when connected to satellite $i$. We consider optical links with circular apertures of radius $r_{iT}$ and $r_{gR}$ for the  transmitter and receiver telescopes at satellite $i$ and ground station $g$ respectively, operating at some wavelength $\lambda$. The free space, and the overall channel transmissivity for such a link is well approximated by, 
\begin{align}
    &\eta_{ig}^{(f)}(t) =\frac{(\pi r_{iT}^2)(\pi r_{gR}^2)}{(\lambda s_{ig}(t))^2},\nonumber\\
    &\eta_{ig}(t) =\eta_{ig}^{(f)}(t)\eta_{ig}^{(a)}(t)
    \eta_T\eta_R,
\end{align}
\noindent where $\eta_{ig}^{(a)}(t), \eta_T,$ and $\eta_R$ are the atmospheric transmissivity, and transmitter and receiver efficiencies respectively.

In general, transmission of a pure state through a pure loss channel reduces the mean photon number of the transmitted state, and produces a mixed state. Hence, this affects the probability that both qubits of the entangled pairs reach both parties successfully. Additionally, since the output state has two-pair terms, loss may cause these terms to lose a single photon and resemble the entangled pair for each party. Note, that such terms are not necessarily entangled, and hence result in a degradation of the fidelity (to the ideal Bell pair) of a distributed pair. Details of the model and the mathematical analysis of the state is given in~\cite{Dhara2021-ch}.

\subsection{Channel Noise}\label{sub:noise}
Additionally, our analysis takes into account excess noise in the optical links (in the form of unfiltered background photons) and dark clicks  in the detectors of the associated hardware. The inclusion of background photons occurs at random and can only be detected when the users detect their individual states. However since the output quantum state is a superposition of multiple terms, the user will be unable to distinguish a background photon from photons generated by the source. As a simple example, in the absence of loss, the detection of the state $\ket{0,1}$ by a user can either mean they received a qubit (of the generated ebit), or they detected the vacuum state with one excess photon (which is not entangled). Hence given a detection pattern, the output state description is best expressed as a statistical mixture of the possible states that could have given rise to the pattern. This limits the fidelity of the output state. Paired with a lossy optical transmission, such a noise process can be highly detrimental to the quality of the shared entanglement. This is simply because, with higher loss there are fewer qubits that are successfully transmitted. Since noise is independent of loss, there is a critical level of background photons above which they significantly eclipse the entangled photons and drive the fidelity of the distributed state below a manageable/suitable threshold. In the present work, we take into account this effect implicitly, by working in loss and noise regimes that remain above the desired fidelity threshold. A detailed analysis of this type of noise is given in Appendix D of  Ref.~\cite{Dhara2021-ch}.

\begin{table}[!htbp]
\begin{center}
		\begin{tabular}{ l|l} 
			\hline
			$S$&Set of satellites\\
			$G$&Set of ground stations\\
			$F$&Set of pair of ground stations\\
			$N$&Number of satellites\\
			$M$&Number of pairs of ground stations\\
			$R_g$&Maximum number of receivers at ground station $g$\\
			$T_i$&Maximum number of transmitters at satellite $i$\\
			$L_j$&Maximum number of simultaneous connections allowed \\
			&for ground station pair $j$\\
			$\theta_e$&Elevation angle limit (from the horizon)\\
			$\kappa$& Total number of time slots\\
			$\Delta$&Slot duration (in seconds)\\
			$t$& A time slot ($1,\cdots, \kappa$)\\
			$e_{ig}(t)$& Elevation angle of satellite $i$ from ground station\\
			&$g$ at time $t$\\
			$F^{th}$&Fidelity threshold\\
			$x_{ij}(t)$&Optimization variable denoting number of connections\\
			&between satellite $i$ and ground station pair $j$ at time $t$\\
			$\omega_{ij}(t)$&Weight associated with the assignment of satellite $i$\\
			&to ground station pair $j$ at time $t$\\
			$\psi_{ij}(t)$ & Entanglement distribution rate associated with satellite $i$ for \\
			&entanglement of the ground station pair $j$ at time $t$\\
			$\chi_{ij}(t)$ & Fidelity associated with satellite $i$ for entanglement\\
			&of the ground station pair $j$ at time $t$\\
			\hline
		\end{tabular}
\end{center}		
		\caption{Summary of Notations.}
		\label{table:smrnd}
\end{table}

%% file: optimization.tex
\section{Quantum Satellite Scheduling Problem}\label{sec:assignment}
We now formulate the scheduling problem in satellite based quantum networks. Time is divided into discrete slots of duration $\Delta$ seconds. Let $t$ be an arbitrary slot. Let $x_{ij}(t)$ be a decision variable denoting number of connections between satellite $i$ and ground station pair $j$ at time $t.$ Our  goal is to assign $S$ to $F$ so as to maximize objective $O(t) = \sum_{j\in F}\sum_{i\in S}\omega_{ij}(t)x_{ij}(t).$ Here, $\omega_{ij}(t)$ denotes the weight/utility associated with the assignment of satellite $i$ to ground station pair $j$. The weight $\omega_{ij}(t)$ can be defined in several different ways: for example, as the {\it entanglement distribution rate} (EDR), denoted by $\psi_{ij}(t)$, or as the average request arrival rate for creating entanglement between ground stations in pair $j$ driven by applications such as: QKD, quantum sensing or quantum teleportation. In the remainder of this paper, we assume that $\omega_{ij}(t) = \psi_{ij}(t)$. 

Note that, additional constraints (e.g., minimum fidelity guarantee or visibility) can be enforced as follows.
\begin{equation}\label{eq:connection-var}
    \omega_{ij}(t)=
\begin{cases}
     \psi_{ij}(t),& \chi_{ij}(t) \ge F^{th} \text{ and } e_{ij_1}(t), e_{ij_2}(t) \ge \theta_e ;\\
     0,& \text{otherwise},
\end{cases}
\end{equation}
where, $\chi_{ij}(t)$ denotes the fidelity of the entangled state distributed to pair $j$ from satellite $i$, $F^{th}$ denotes the minimum fidelity threshold, and $e_{ij_1}(t)$ and $e_{ij_2}(t)$ are the elevation angles of satellite $i$ from ground stations $j_1$ and  $j_2$ respectively. Here, $j = \{j_1, j_2\}$, i.e. pair $j$ consists of ground stations $j_1$ and $j_2.$ Note that, $\omega_{ij}(t)$ and  $\chi_{ij}(t)$ can be determined according to the noise and loss models discussed in Section II.B.

The optimal transmission scheduling problem at time slot $t$ can be cast as the following integer programming problem.
\begin{subequations}\label{eq:sat-opt}
\begin{align}
&\text{\bf{\prs:}} \nonumber\\
&\max \quad\sum_{j\in F}\sum_{i\in S} \omega_{ij}(t) x_{ij}(t)\label{eq:stat-sat1}\displaybreak[0]\\
\text{subject to}&\sum_{i\in S}\sum_{j\in F, g\in j} x_{ij}(t) \le R_g, \forall g\in G,\label{eq:stat-sat2}\\
\quad&\sum_{j\in F} x_{ij}(t) \le T_i, \forall i \in S\label{eq:stat-sat3}\\
\quad&\sum_{i\in S} x_{ij}(t) \le L_j, \forall j \in F\label{eq:stat-sat4}\\
&\text{s.t.} \quad x_{ij}(t)\in \{0,1,2,\cdots\} \quad \forall i \in S, \forall j\in F,\label{eq:stat-sat5}
\end{align}
\end{subequations}
\noindent where $T_i$ and $R_g$ are the maximum number of transmitters and receivers at  satellite $i$ and ground station $g$ respectively. $L_j$ denotes the maximum number of simultaneous connections allowed for pair $j$. We summarize the notations used in Table \ref{table:smrnd}. 
\begin{figure}[htbp]
\begin{minipage}{0.3\textwidth}
\includegraphics[width=0.85\textwidth, height=0.45\textwidth]{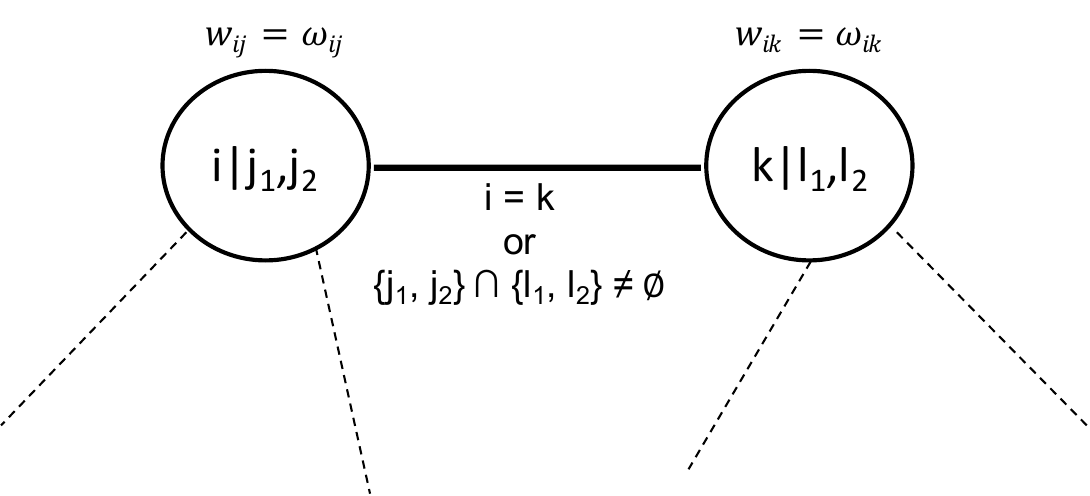}
\subcaption{}
\end{minipage}
\begin{minipage}{0.3\textwidth}
\includegraphics[width=0.5\textwidth, height=0.8\textwidth]{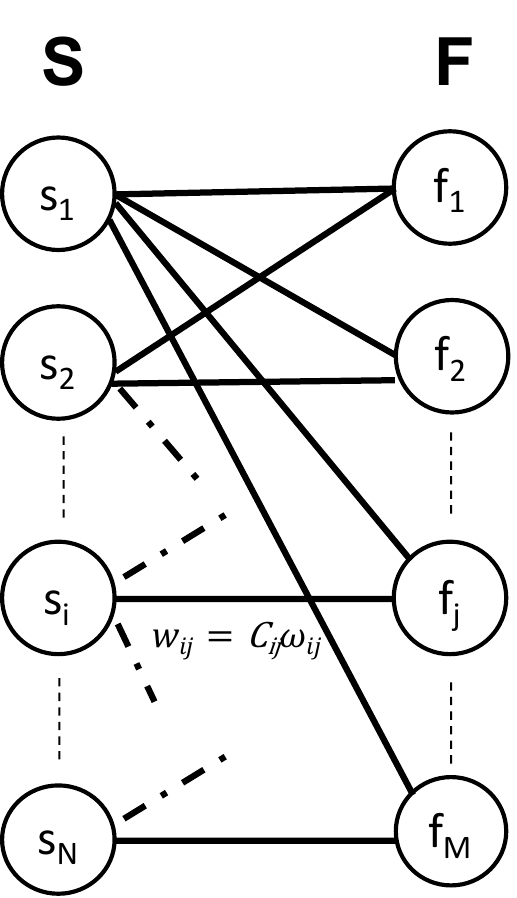}
\subcaption{}
\end{minipage}
\caption{Solving \prs\;can be equivalent to (a) solving a maximum weight independent set problem when $R_g = 1, T_i=1, L_j = 1$ and (b) solving a matching problem when $R_g = N, T_i=1, L_j = 1$.}
\label{fig:graphs}
\end{figure}

A ground station $g$ can be part of multiple ground station pairs and thus is not allowed  to be allocated to more than $R_g$ satellites. This constraint is enforced in \eqref{eq:stat-sat2}. Constraint \eqref{eq:stat-sat3} ensures that satellite $i$ does not get allocated to more than $T_i$ ground station pairs. Similarly, Constraint \eqref{eq:stat-sat4} does not allow ground station pair $j$ to be allocated to more than $L_j$ satellites. We also assume $R_g \ge L_j, \forall j\in F, g \in j$. Constraint \eqref{eq:stat-sat5} ensures that $\{x_{ij}(t)\}$ are integer decision variables.
\subsection{Special Cases}
Below we consider few specific scenarios and in some cases, design efficient exact algorithms to solve \prs. 
\subsubsection{Single transmitter and single receiver scenario - STSR ($R_g = 1, T_i=1, L_j = 1$)}
We first consider the case when each satellite has only a single transmitter and each ground station has a single receiver. We can construct a graph $G_t(V, E)$ as shown in Figure  \ref{fig:graphs}(a). We set $V = \{(i|j_1,j_2), \forall i \in S, j = \{j_1, j_2\} \in F, \omega_{ij}(t) >0\}.$ We assign each vertex $v \in V$ a weight $w_v = \omega_{ij}(t)$ where $v = (i|j_1,j_2)$ and $j = \{j_1, j_2\}.$ Also we define $E = \{((i|j_1,j_2), (k|l_1,l_2)) \text{ s.t. } i=k \text{ or } \{j_1,j_2\} \cap \{l_1,l_2\} \ne \phi\}.$ It can be easily seen that solving \prs\;in this case corresponds to obtaining the maximum weight independent set of $G_t.$ However, this problem is known to be NP-hard for general graph . Numerous approximation algorithms have been developed to solve the maximum weight independent set problem.

\subsubsection{Single transmitter and multiple receiver scenario - STMR ($R_g = N, T_i=1, L_j = 1$)}
Next we consider the case when each satellite has only a single transmitter and each ground station has multiple receivers. If we are restricted to assign only one satellite to one ground station pair, then solving \prs\;can be transformed into finding maximum weight matching in a bipartite graph $G_t(V, E)$ with bi-partition $(S, F)$ and $V = S\cup F$ as shown in Figure \ref{fig:graphs} (b). In this case, we define a weight function on the edges $w: E \rightarrow \mathbb{R}$ as follows. We set $w_{ij} = \omega_{ij}(t)$ for all $i \in S, j \in F.$ The maximum weight bipartite matching problem on $G_t$ can be solved efficiently by the Hungarian algorithm \cite{kuhn1955hungarian} with a time complexity of $O(n^3)$ where $n = \max\{N, M\}$.

Note that, the case with $R_g = N, T_i\ge 1, L_j \ge 1$ can be efficiently solved by creating $T_i$ and $L_j$ copies of satellite $i$ and ground station pair $j$ respectively at prescribed locations and by finding the optimal bipartite matching on this extended bipartite graph.

\subsection{Fairness}
The solution of \prs\;in \eqref{eq:sat-opt} maximizes the aggregate EDR, ignoring the EDR achieved by each ground station pair. This may result in an unfair allocation with some ground station pairs getting higher individual EDRs compared to others. In particular, ground station pairs that are farther apart or located in less favorable conditions (e.g., in inclement weather conditions) may receive significantly lower individual EDRs compared to what they could have achieved otherwise. To address this, we introduce the \prf\;algorithm based on the principles of iterative weighted max-min fairness \cite{Zehavi2013}.

The main idea behind \prf\;is to iteratively solve a max-min optimization problem. Instead of maximizing the total EDR, the objective is to ensure that the minimum EDR received by any ground station pair is maximized. To formalize this, we define the following ONE-SHOT-MAXMIN problem, which seeks to maximize the minimum weighted EDR across all ground station pairs in a single optimization step:
\begin{align}\label{max-min-iter}
&\text{\bf{ONE-SHOT-MAXMIN:}} \nonumber\\
&\max \bigg[\min_{j\in F} \bigg(\sum_{i\in S} f_{ij}(t) x_{ij}(t)\bigg)\bigg]\displaybreak[0]\nonumber\\
&\text{subject to: }\text{constraints }\eqref{eq:stat-sat2}, \eqref{eq:stat-sat3}, \eqref{eq:stat-sat4}, \eqref{eq:stat-sat5},
\end{align}
\noindent where, $f_{ij}(t)$ denotes the fairness adjusted EDR weight for the connection between satellite $i$ and ground station pair $j.$ 

The weights $f_{ij}(t)$ may be defined based on the desired fairness model: such as $f_{ij}(t) = \omega_{ij}(t)$ for absolute EDR. Alternatively, we compute the highest achievable EDR for each ground station pair $j$ assuming that the entire satellite network's resources are available to exclusively serve that pair. We denote this maximum possible EDR as $A_j(t).$ We set $f_{ij}(t) = \omega_{ij}(t)/A_j(t)$ and refer to this ratio as fractional (normalized) EDR.
\begin{algorithm}
\caption{A weighted fair scheduling algorithm for quantum satellite networks}\label{algo-maxmin}
\begin{algorithmic}[1]
\Procedure{\prf}{}
\State Let $\bar{X} =[\bar{x}_{ij}(t), i\in S, j\in F]$ be the  optimal solution of the ONE-SHOT-MAXMIN problem defined in \eqref{max-min-iter}. 
\State Let $X^* = \bar{X};$ allocated\_pair $=\phi$. 
\While{$F \ne \phi$} \label{line-end}
\State $\bar{f}_j = \sum_{i\in S} f_{ij}(t) \bar{x}_{ij}(t), j\in F$
\State $F' = arg\,min\{\bar{f}_j, j\in F\}$.
\For{each pair $j$ in $F'$}
\State allocated\_pair $\leftarrow$ allocated\_pair $\cup \{j\}$
\For{each satellite $i$ in $S$}
\State $T_i \leftarrow T_i - \bar{x}_{ij}(t)$ 
\State $R_g \leftarrow R_g - \bar{x}_{ij}(t), \forall g \in j$
\State Remove satellite $i$ from $S$ if $T_i=0$.
\State Remove GS $g$ from $G$ if $R_g=0$.
\EndFor
\State $F = F\setminus \{j\}.$
\EndFor
\State Let $\bar{X}$ be the optimal solution of the ONE-SHOT-MAXMIN problem defined in \eqref{max-min-iter} with the new values of $T_i, R_g, S,$ and $F.$
\If{$\bar{X} = \phi$}\label{line_feasibility}
\Return $X^*$
\Else{}
\For{each pair $j^{'}$ in $F$}
\For{each satellite $i$ in S}
\State $X^{*}[i,j^{'}] = \bar{X}[i,j^{'}]$
\EndFor
\EndFor
\EndIf
\EndWhile
\State \Return $X^*$
\EndProcedure
\end{algorithmic}
\end{algorithm}

We present \prf\; in Algorithm \ref{algo-maxmin} and adopt an iterative optimization approach to obtain a fair allocation $X^*.$ In the first iteration, we solve the linear program\footnote{The optimization problem \eqref{max-min-iter} can be transformed into a linear program by introducing a variable $\lambda$. The goal would then be to maximize $\lambda$ while ensuring resource constraints and that every pair's weighted EDR in the optimal allocation is at least $\lambda.$} \eqref{max-min-iter} to find an allocation that maximizes the lowest weighted EDR. Next, we determine the pairs that cannot exceed this allocation and fix their assigned values. We then iteratively maximize the next lowest allocation until either all allocations are fixed or we no longer have resources to generate a feasible solution (Lines \ref{line-end} and \ref{line_feasibility} in Algorithm \ref{algo-maxmin}).

\begin{remark}
    Our simulation results show that assigning $f_{ij}(t)$ based on fractional EDRs produce more fair allocations than absolute EDR weights.
\end{remark}

%% file: reflector.tex
\section{Quantum Satellite Scheduling Problem: Reflection Based Formulation}\label{sec:reflection}
\begin{figure}[!htbp]
\centering
\begin{minipage}{0.5\textwidth}
\includegraphics[width=1\textwidth]{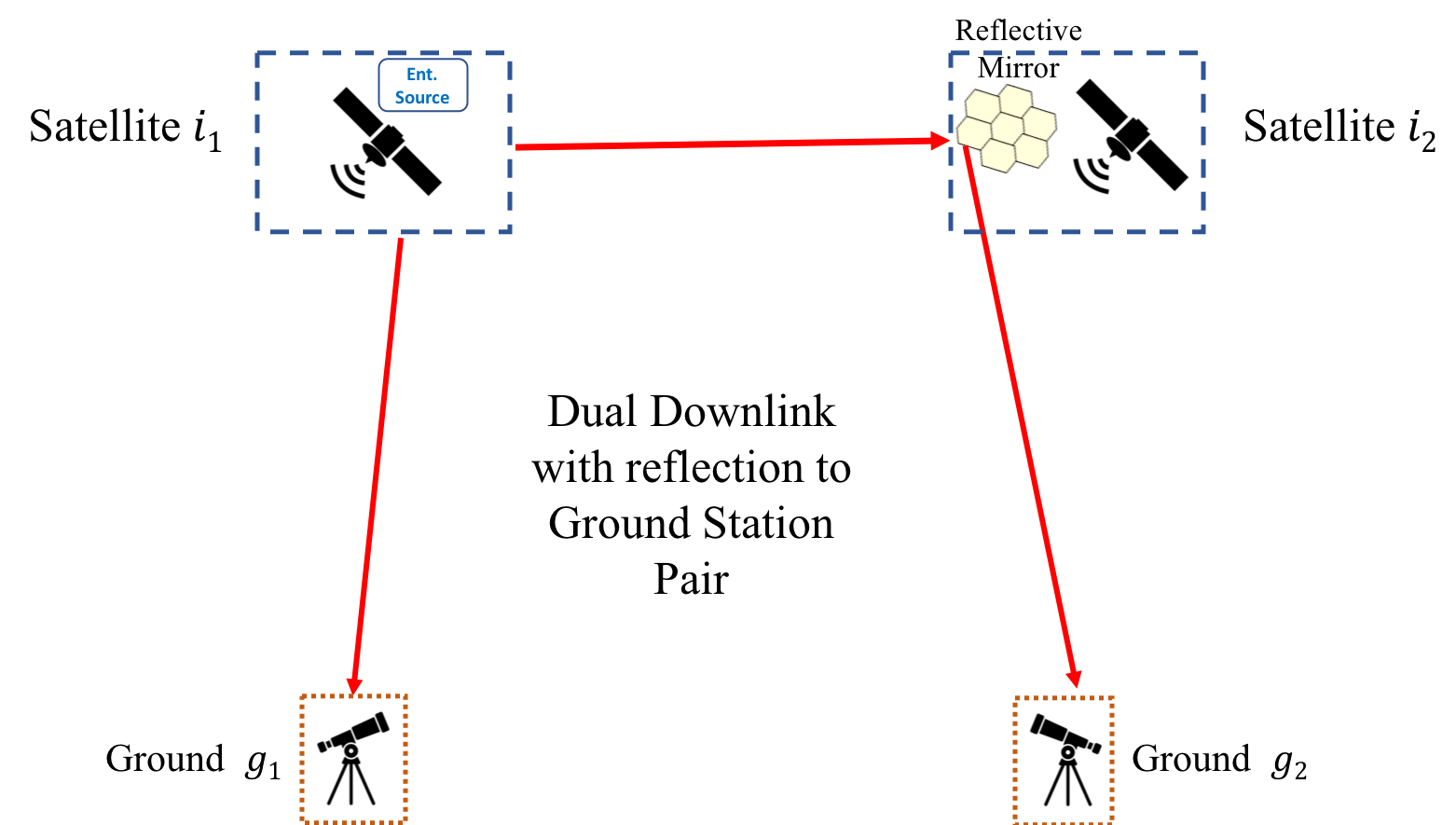}
\end{minipage}\hfill
\caption{Dual downlink architecture for Reflection Based Scheme. This architecture involves two satellites: a primary satellite equipped with an entangled photon source and a secondary reflecting satellite. The primary satellite generates a pair of entangled photons, directing one photon to a ground station and the other to the reflecting satellite. The reflecting satellite then forwards the received photon to a second ground station, effectively establishing a dual downlink for entanglement distribution.}
\label{dualdownlink-reflection-new}
\end{figure}

\begin{figure}[ht]
    \centering
    \begin{subfigure}{0.49\columnwidth}
        \centering
        \includegraphics[width=\linewidth]{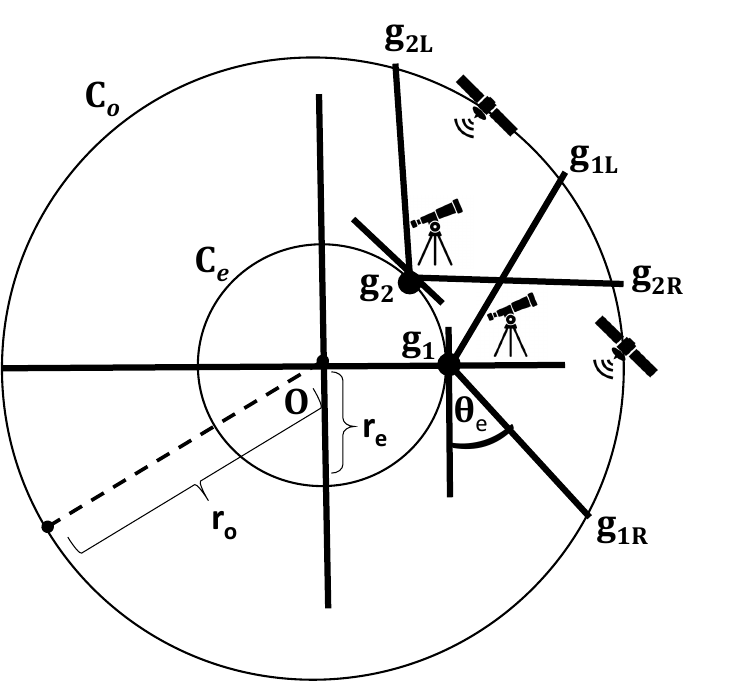} 
        \caption{}
        \label{fig:rate-fid-Zero-Pd}
    \end{subfigure}
    \hfill
    \begin{subfigure}{0.49\columnwidth}
        \centering
        \includegraphics[width=\linewidth]{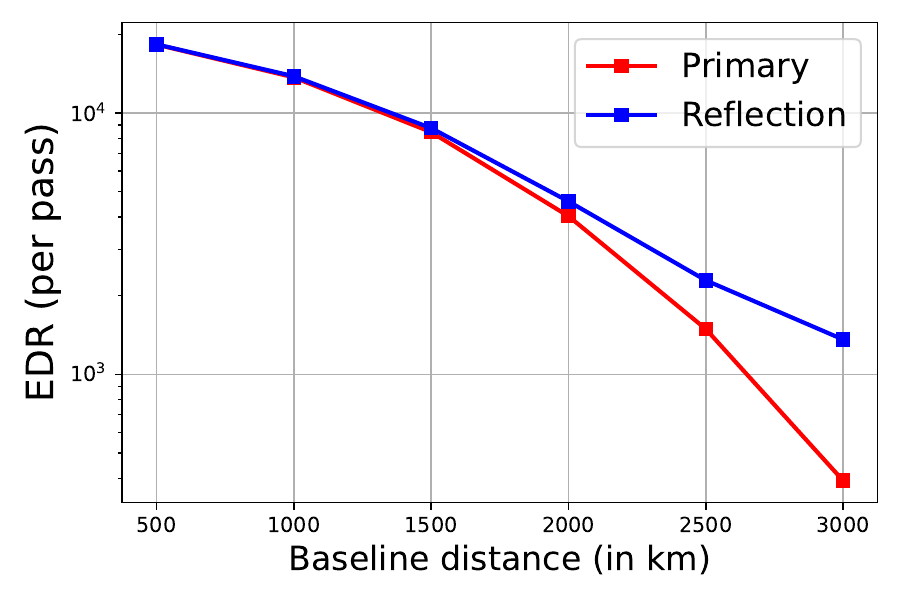}
        \caption{}
        \label{fig:rate-fid-high-Pd}
    \end{subfigure}
    \caption{Comparison of Primary versus Reflection based entanglement distribution schemes.}
    \label{primary-vs-reflection}
\end{figure}

The path traversed by the entangled pair in vacuum  experiences fewer losses and decoherence. Thus one can take advantage of using two satellites \cite{goswami2023satellite} to serve entangled pairs to ground stations: a source satellite and a reflecting satellite as shown in Figure \ref{dualdownlink-reflection-new}. The source satellite generates entangled photon pairs, sending one to a ground station and the other to the reflecting satellite which has a mirror that reflects the incident photon towards the other ground station. It should be emphasized that no quantum memories are needed for this scheme. We assume the reflection at the mirror of the second satellite does not impact the fidelity of the entangled pair, though it may lead to some loss.

In Figure \ref{primary-vs-reflection}, we illustrate a case study conducted on a simplified setup consisting of two satellites and a single pair of ground stations. The goal of this study is to identify the parameter regimes where a reflection based scheme would provide significant benefits. We present the set-up in Figure \ref{primary-vs-reflection} (a). Let $g_i$ denote the location of ground station $i$ on earth for $i = 1, 2$. The great circle $C_e$ connects $g_1$ and $g_2,$ while $C_o$ represents the orbital path of the satellite. Denote $r_e$ and $r_o$ as the radii of the earth and the satellite orbit, respectively. 

We consider satellites moving in an overhead orbit\footnote{This allows us to obtain closed form expressions for satellite-ground station distances using trigonometric identities.}, i.e., the satellite orbit and the great circle lie in the same plane. Let $\theta_e$ denote the minimum elevation angle (from horizon) required for satellite visibility. Let $g_{iL}$ and $g_{iR}$ represent the positions on the orbit at which a satellite appears at elevation $\theta_e$ from ground station $i$, viewed in the counterclockwise and clockwise direction, respectively. A satellite $i$ is visible to ground station $i$ while it traverses the arc $(g_{iL}, g_{iR}).$ Thus, simultaneous visibility to both ground stations and hence an implementation of the primary based scheme, is only possible when a satellite is on the overlapping arc $(g_{1L}, g_{2R}).$ However, a reflection based scheme broadens the range and enables entanglement distribution even when one satellite is on $(g_{2L}, g_{1L})$ and the other is on $(g_{2R}, g_{1R}).$

We plot the EDR per orbital pass as a function of the geodesic (baseline) distance between $g_1$ and $g_2$ in Figure \ref{primary-vs-reflection} (b). When the distance between ground stations is small, the visibility arcs from both ground stations overlap substantially, limiting the benefit of the reflection based scheme. However, as the baseline increases, the overlap diminishes and the potential gain from reflection becomes evident. At a baseline distance of $3000 KM$, we observe nearly a three fold increase in EDR for the reflection based scheme compared to the primary scheme. While this demonstrates improvement for a single satellite and ground station pair, we extend this analysis in Section \ref{sec:numeric} to evaluate reflection based gains across multiple ground station pairs and a satellite constellation over a full day of simulation.

Building on the insights from the two-satellite case study, we now generalize our approach to a constellation level setting involving multiple satellites and ground station pairs. We cast the scheduling problem as an Integer Linear Program as shown in \eqref{eq:reflection-opt}, that considers both primary and reflection based entanglement distribution options. 
\begin{subequations}\label{eq:reflection-opt}
\begin{align}
&\text{\bf{\rrs:}}\nonumber\\ 
&\max \sum_{i\in S}\sum_{j\in F} \omega_{ij}(t) x_{ij}(t) 
+ \sum_{i\in S}\sum_{k\in S}\sum_{j\in F} \nu_{ikj}(t)y_{ikj}(t)\label{eq:rf-stat-sat1}\displaybreak[0]\\
&\text{subject to:}\nonumber\\ 
&\sum_{i\in S}\sum_{j\in F, g\in j} x_{ij}(t) + \sum_{i\in S}\sum_{k\in S}\sum_{j\in F, g\in j} y_{ikj}(t) \le R_g, \forall g\in G,\label{eq:rf-stat-sat2}\\
&\sum_{j\in F} x_{ij}(t) +  \sum_{j\in F}\sum_{k\in S}y_{ikj}(t) \le T_i, \forall i \in S\label{eq:rf-stat-sat3}\\
&\sum_{j\in F}\sum_{i\in S}y_{ikj}(t) \le U_k, \forall k \in S \label{eq:rf-stat-sat4}\\
&\sum_{i\in S} x_{ij}(t) + \sum_{i\in S}\sum_{k\in S} y_{ikj}(t) \le L_j, \forall j \in F \label{eq:rf-stat-sat5}\\
&\text{s.t.} \quad x_{ij}(t), y_{ikj}(t) \in \{0,1,2,\cdots\}, \quad \forall i, k \in S, \forall j\in F,\label{eq:rf-stat-sat6}
\end{align}
\end{subequations}

The objective of \rrs\;is to maximize the aggregate EDR across all ground station pairs, accounting for both primary and reflection based entanglement distributions. To model the reflection based set-up, we introduce the variable $y_{ikj}(t),$ which denotes the number of reflection assisted connections involving satellite pair $(i, k)$ and ground station pair $j$. We assume an implicit ordering, i.e. satellite $i$ serves the first ground station in pair $j$ and satellite $k$ serves the second. We denote $\nu_{ikj}(t)$ as the EDR achieved when satellite pair $(i, k)$ is assigned to ground station pair $j$. We also capture the feasibility of a reflection based connection, i.e. $\nu_{ikj}(t) > 0$  only if satellite $i$ is visible to the first ground station, satellite $k$ is visible to the second and the two satellites are mutually visible to each other. 

In addition to the resource constraints captured in the \prs\;formulation, we introduce an additional constraint in \eqref{eq:rf-stat-sat4}, specific to the reflection based set-up. We denote $U_k$ as the number of reflectors (or mirrors) on satellite $k$ and pose \eqref{eq:rf-stat-sat4} as a constraint, limiting the number of simultaneous reflection based connections that satellite $k$ can support.

\subsection{Fairness}
We also design an iterative max-min based fair resource allocation algorithm for the reflection based set-up and present its details in Algorithm \ref{algo-maxmin-ref}.
\begin{algorithm}
\caption{A weighted fair scheduling algorithm for Reflection based set-up}\label{algo-maxmin-ref}
\begin{algorithmic}[1]
\Procedure{\rrf}{}
\State Let $\bar{X}, \bar{Y}=[\bar{x}_{ij}(t), i\in S, j\in F], [\bar{y}_{ikj}(t), i, k \in S, j\in F]$ be the optimal solution of the following optimization problem. Define $A_j(t)$ as the maximum possible EDR achievable for pair $j$, when there is no contention for resources.
\begin{align}\label{ref-max-min-iter}
&\max \bigg[ \min_{j\in F} \bigg(\sum_{i\in S}\omega_{ij}(t)x_{ij}(t) \nonumber \\
&\quad\quad\quad\quad\quad+ \sum_{i\in S}\sum_{k\in S} \nu_{ikj}(t)y_{ikj}(t)\bigg)/A_j(t)\bigg]\nonumber\\
&\text{subject to: }\text{constraints }\eqref{eq:rf-stat-sat2}, \eqref{eq:rf-stat-sat3}, \eqref{eq:rf-stat-sat4}, \eqref{eq:rf-stat-sat5}, \eqref{eq:rf-stat-sat6}
\end{align}
\State  Let $X^*, Y^* = \bar{X}, \bar{Y}$.
\While{$F \ne \phi$} \label{ref-line-end}
\State Find GS pairs that have saturated.
\State Remove saturated pairs and update $F$.
\State Adjust available resources to reflect those consumed by saturated pairs. 
\State Recompute $\bar{X}, \bar{Y}$ as the optimal solution of \eqref{ref-max-min-iter} with the new values of resources and $F.$
\State Update $X^*$ and $Y^*$.
\EndWhile
\State \Return $X^*$ and $Y^*$.
\EndProcedure
\end{algorithmic}
\end{algorithm}
The \rrf\;algorithm builds on the same principles as the \prf\;algorithm (Algorithm \ref{algo-maxmin}). As in \prf\;, \rrf\;begins by solving a max-min ILP that maximizes the minimum fractional EDR across all ground station pairs subject to various resource constraints. It then identifies saturated ground station pairs whose EDRs can not be increased further and removes them from further consideration. The algorithm then again solves the maxmin ILP over the remaining ground station pairs and updated set of resources in the next round of optimization. This process continues until no further EDR improvements are possible for any pair.

%% file: simulations.tex
\section{Performance Evaluation}\label{sec:numeric}

\begin{figure*}[!htbp]
\centering
\hspace{-0.5cm}
\begin{minipage}{0.5\textwidth}
\includegraphics[width=0.83\textwidth]{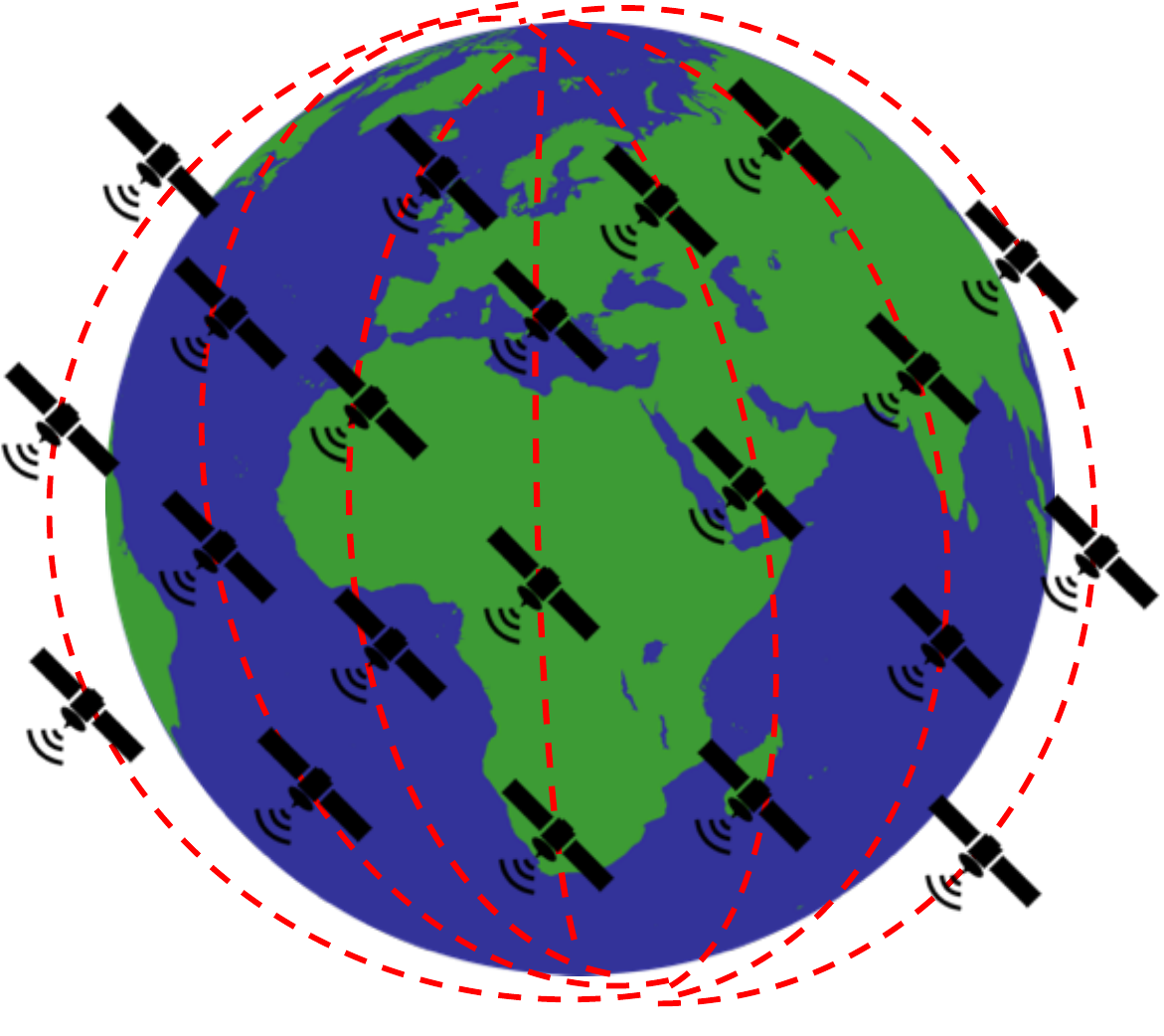}
\vspace{0.8cm}
\subcaption{}
\end{minipage}
\hspace{-1.5cm}
\begin{minipage}{0.5\textwidth}
\includegraphics[width=1.1\textwidth, height = 0.8\textwidth]{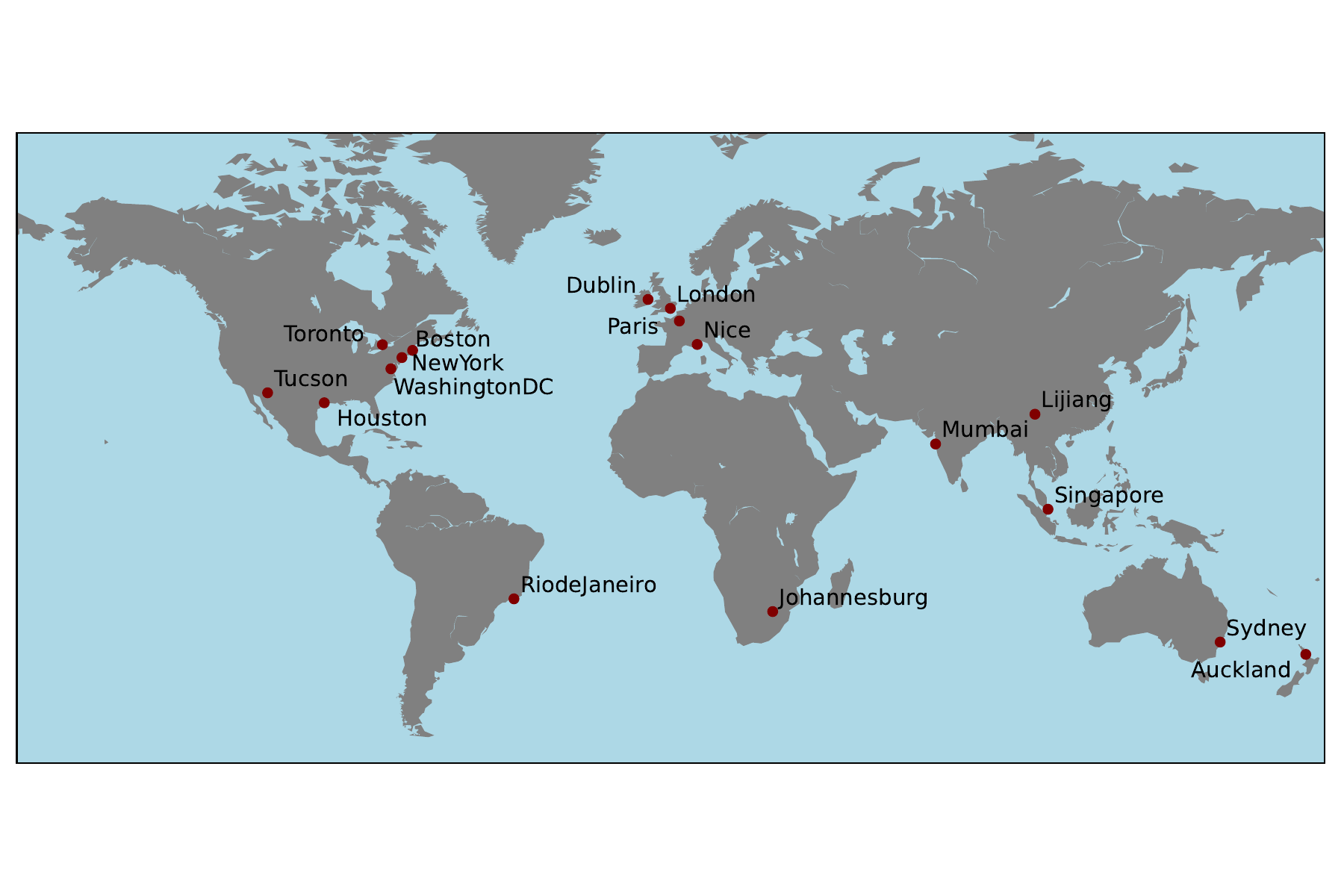}
\subcaption{}
\end{minipage}
\caption{Orbital and ground station layout.
(a) Satellite constellation with Polar orbits: illustration of the satellite network, where satellites follow polar orbital paths to ensure global coverage, enabling repeated passes over ground stations at different latitudes. (b) Ground station locations: geographic distribution of ground stations across the Earth's surface, selected to support regional and global entanglement distribution.}
\label{sat-gs}
\end{figure*}

\begin{figure*}[!htbp]
\centering
\hspace{-1cm}
\begin{minipage}{0.3\textwidth}
\includegraphics[width=1\textwidth]{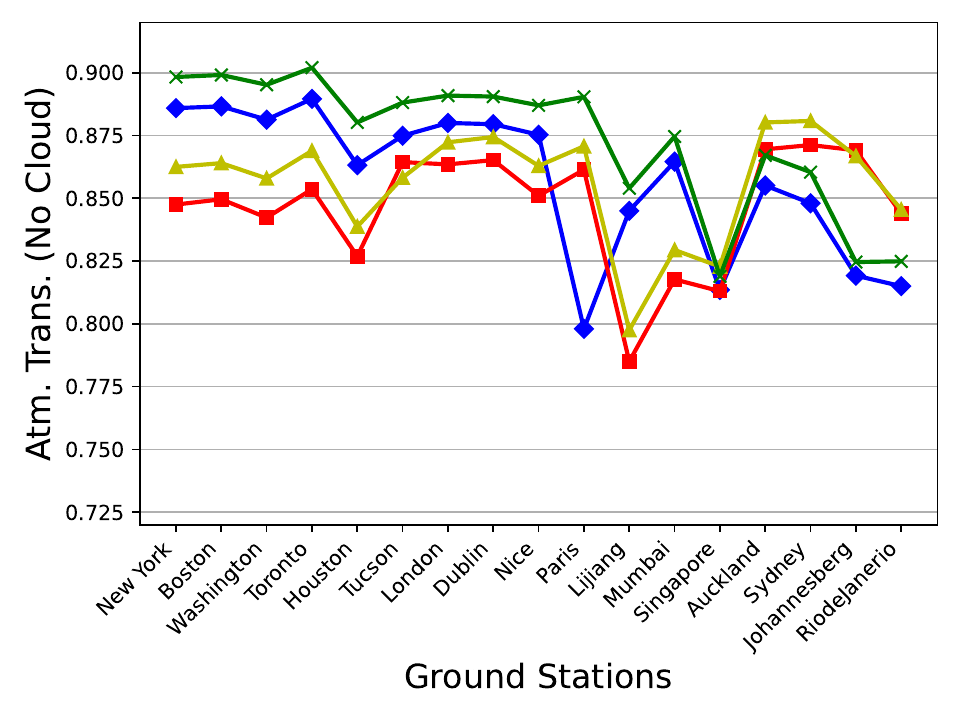}
\subcaption{}
\end{minipage}
\begin{minipage}{0.3\textwidth}
\includegraphics[width=1\textwidth, height = 0.75\textwidth]{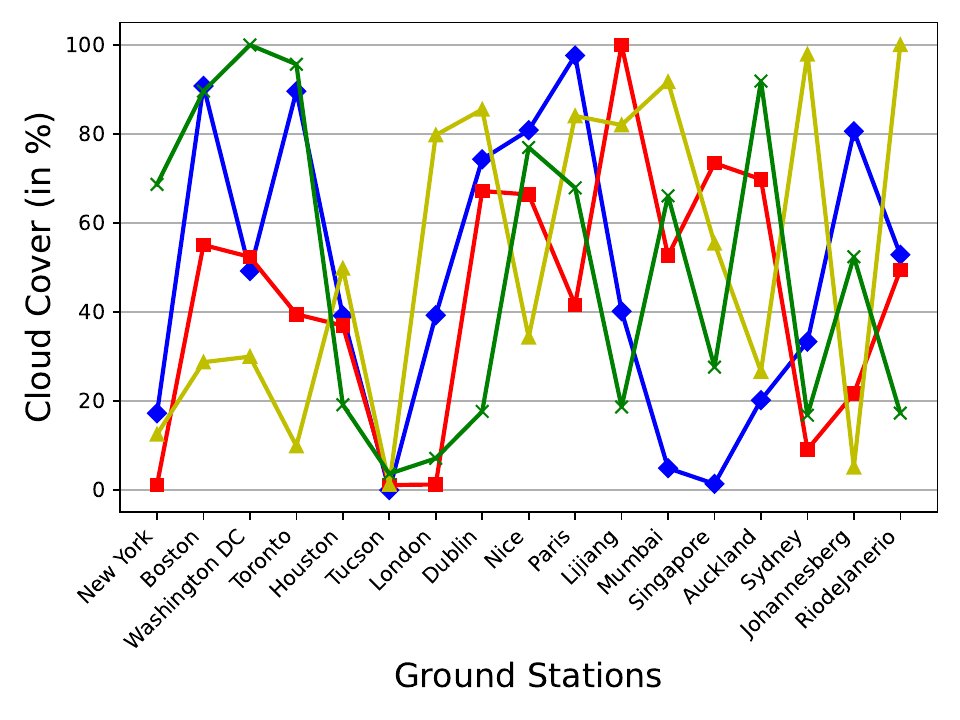}
\subcaption{}
\end{minipage}
\begin{minipage}{0.3\textwidth}
\includegraphics[width=1.2\textwidth, height = 0.8\textwidth]{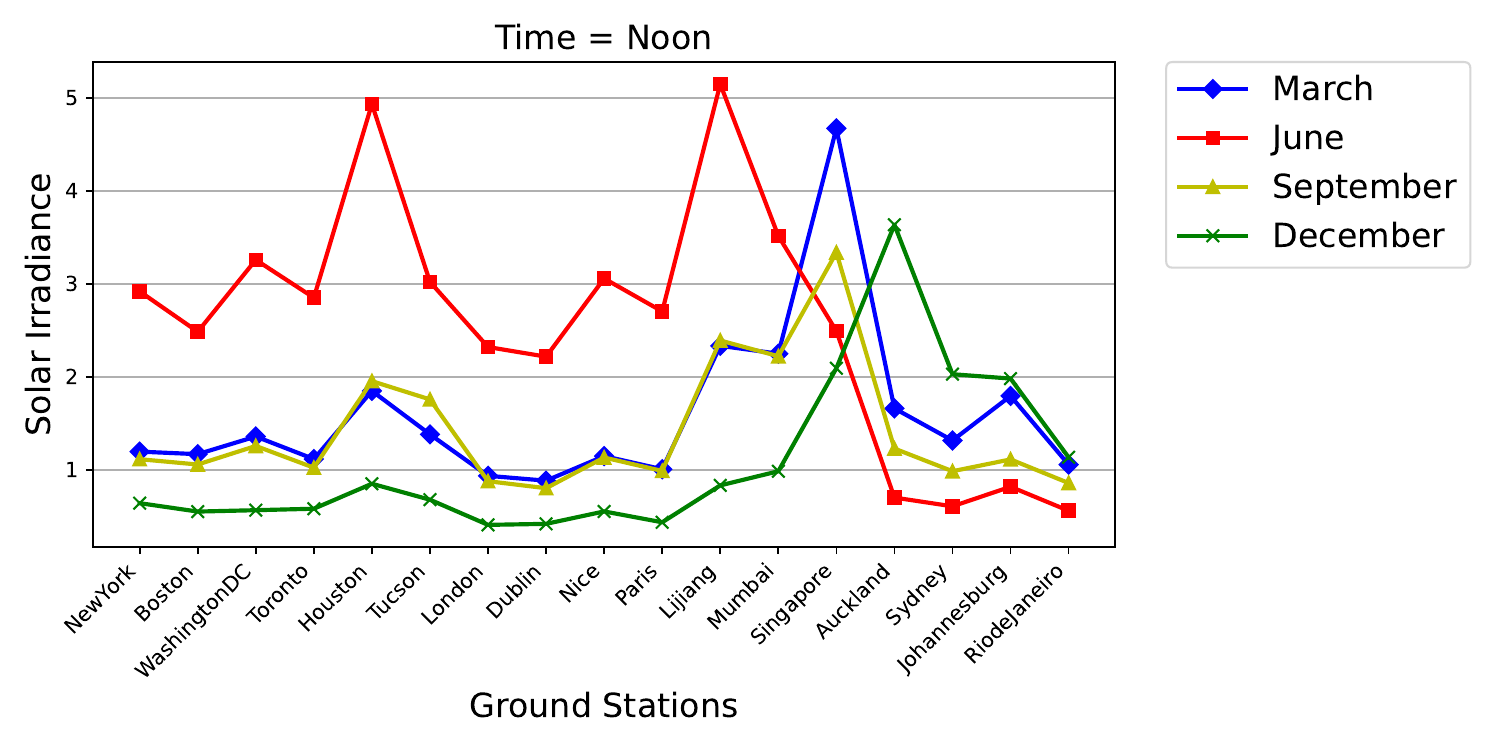}
\subcaption{}
\end{minipage}
\caption{Weather-related parameters affecting entanglement distribution.
(a) Atmospheric Transmissivity: represents the fraction of optical photons that successfully passes through the atmosphere, influenced by aerosols, humidity, and atmospheric composition. (b) Cloud Coverage: indicates the proportion of sky obscured by clouds, which directly impacts line-of-sight visibility and link availability for satellite-to-ground entanglement distribution. (c) Solar Irradiance: measures the intensity of sunlight received at the Earth's surface, which contributes to background photon noise and can affect the signal-to-noise ratio at ground-based quantum receivers.}
\label{weather}
\end{figure*}
\begin{table}[!htbp]
\begin{center}
		\begin{tabular}{ |l|l|l|l|l|l|} 
			\hline
                Param&Value&Param&Value&Param&Value\\
                \hline
			$\theta_e$&$20^{\circ}$&$\Delta$&$10$ s&$\kappa$&$86400$ s\\
			\hline
                $F^{th}$&0.85&$r_{iT}$&$0.1$ m&$r_{gR}$&$1$ m\\
                \hline
                $\lambda$&$737$ nm& Rep. rate&$10^9$&$N_S$&$0.0078$\\
                \hline
                $\eta_T$&$0.7$&$\eta_R$&$0.7$&$T_i,R_g, L_j, U_k$&$10$\\
                \hline
		\end{tabular}
\end{center}		
		\caption{Summary of simulation parameters.}
		\label{table:sim-params}
\end{table}
We evaluate the effectiveness of the proposed scheduling policies using a polar satellite constellation where satellites are deployed in $20$ equally spaced rings, each consisting of $20$ equally spaced satellites at the same orbit altitudes (See Figure \ref{sat-gs}a). More details on this constellation can be found in \cite{Khatri21}. We assume the ground station pairs to be all possible pair combinations of $17$ major cities across the globe as shown in Figure \ref{sat-gs}b. We summarize the simulation parameters used in our numerical experiments in Table \ref{table:sim-params}. We consider an entanglement source with a repetition rate of $1GHz$, i.e., $10^{9}$ entanglements are generated at the source per second within each time slot. 

\subsection{Weather and Time of day effect modeling}\label{sub-trans}
We next incorporate models for weather and time of day to capture the impact of seasonal and environmental conditions on entanglement distribution.

In a quantum satellite network, downlink entanglement distribution occurs over a free space channel between satellite and ground station. However, atmospheric effects can greatly reduce its transmission efficiency. To quantify this effect, we employ MODTRAN\footnote{MODTRAN is a widely used radiative transfer model covering wavelengths from the ultraviolet to the thermal infrared.}, a widely used atmosphere modeling tool. MODTRAN models the atmosphere as a sequence of homogeneous layers and calculates absorption coefficients at standard pressure and temperature, using spectroscopic data from its built-in database \cite{MODTRAN5}. The Atmosphere Generator Toolkit (AGT) adds to MODTRAN by creating custom atmospheric profiles from historical and radiosonde data. It provides long-term, time-averaged atmospheric conditions of specific geographic locations using their latitude and longitude information. 

For our analysis, we use MODTRAN to compute atmospheric transmissivity ($\hat{\eta}_{g}^{(a)}(t)$) at zenith under clear sky conditions, assuming full visibility and no cloud cover. We denote $\theta_{ig}(t)$ as the angle of elevation between satellite $i$ and ground station $g$ at time $t.$ We model the relationship between actual atmospheric transmissivity and the angle of elevation as \cite{dequal2021feasibility}:
\begin{align}
    \eta_{ig}^{(a)}(t) = [\hat{\eta}_{g}^{(a)}(t)]^{\text{sec}(90-\theta_{ig}(t))}
\end{align}

\begin{figure*}[htbp]
\centering
\hspace{-1cm}
\begin{minipage}{0.5\textwidth}
\includegraphics[height = 1.2\textwidth, width=1\textwidth]{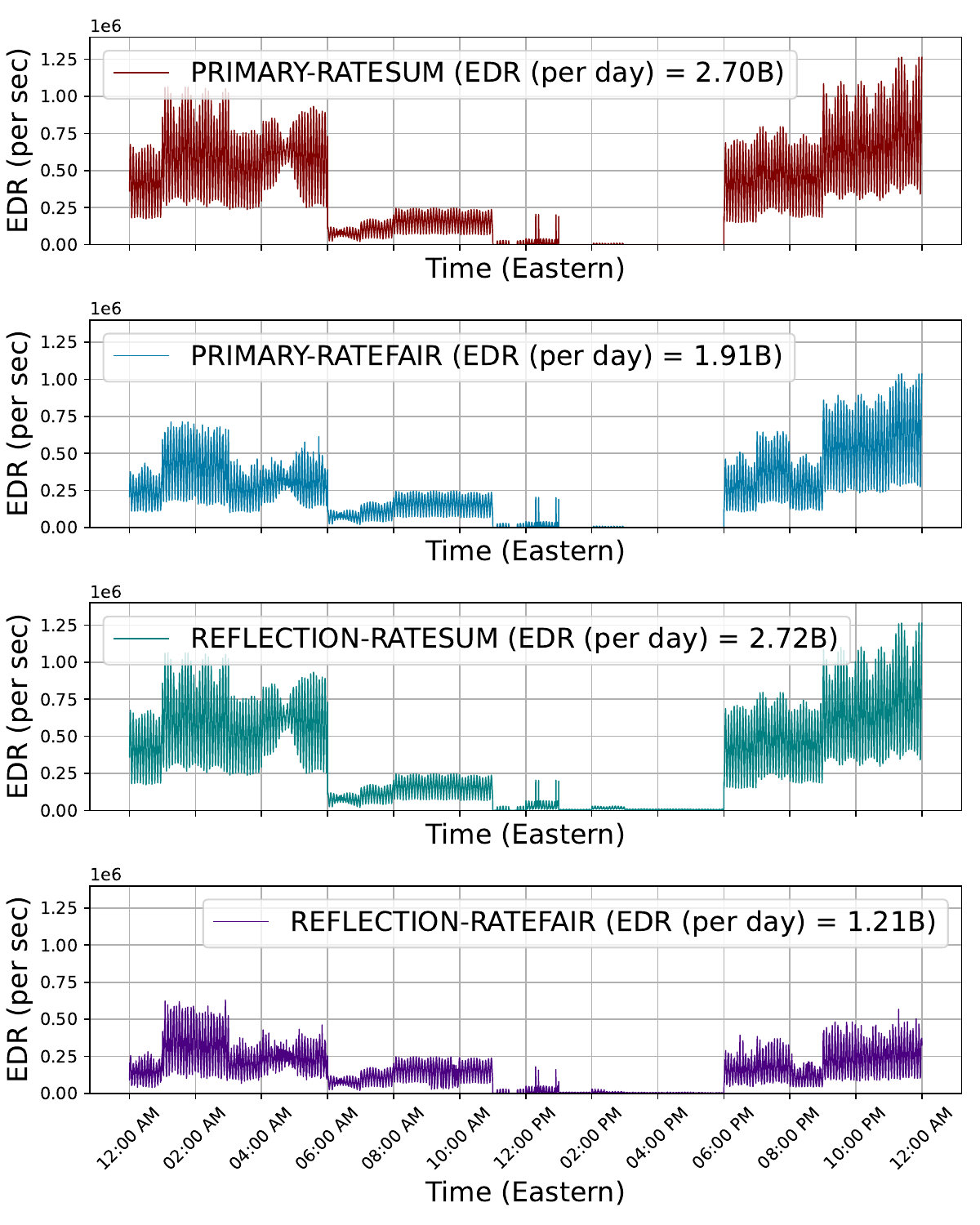}
\subcaption{}
\end{minipage}
\hspace{-0.3cm}
\begin{minipage}{0.5\textwidth}
\includegraphics[height = 1.1\textwidth, width=1\textwidth]{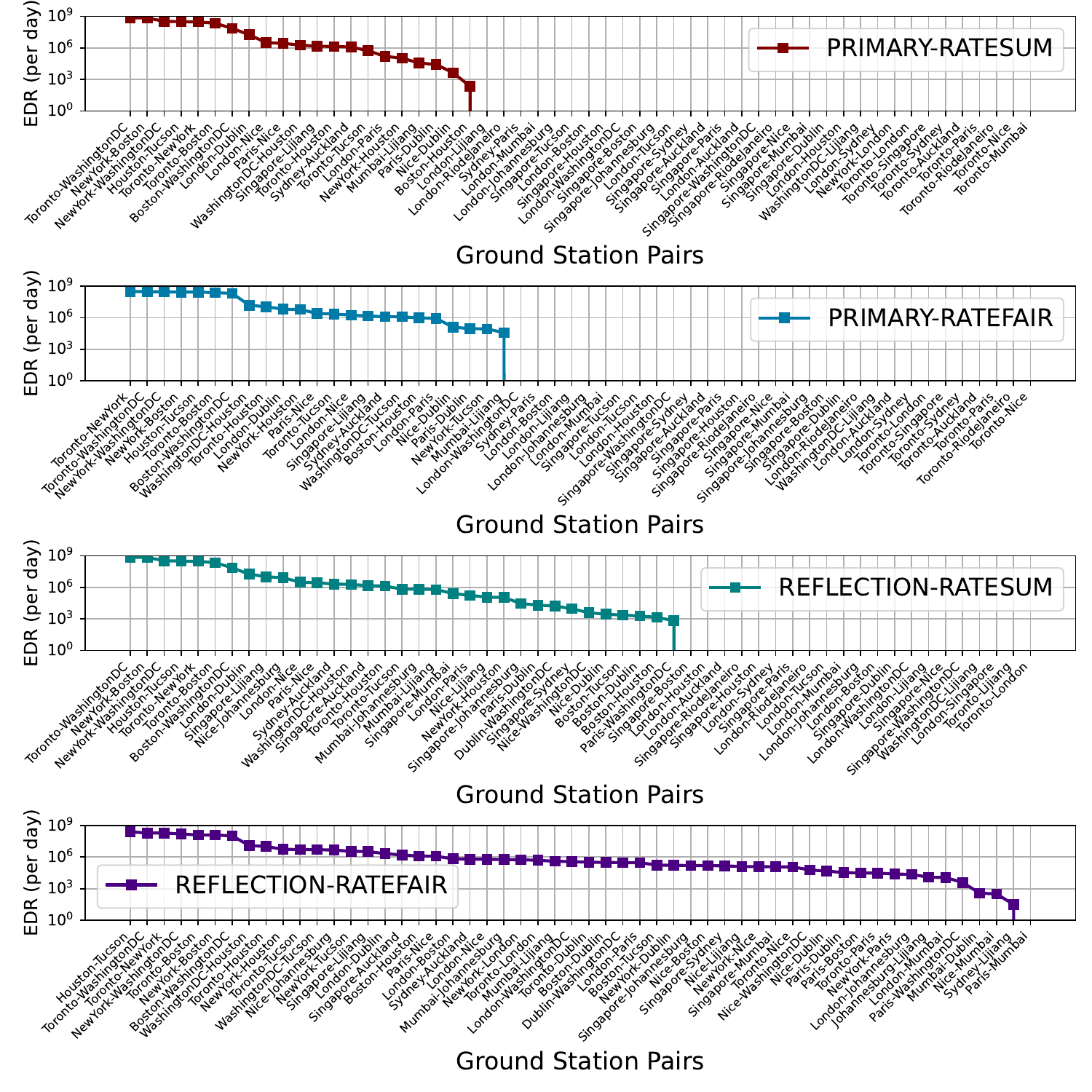}
\subcaption{}
\end{minipage}\hfill
\caption{Performance comparison of four scheduling policies. (a) aggregated EDR over time across all ground station pairs. (b) Per day EDR for individual ground stations. The satellite orbit altitude is set to 1000 km, with $T = R = L = 10$ ($T, R, L$ denote the maximum number of pair assignments per satellite, satellite assignments per ground station, and satellite assignments per ground station pair, respectively) and simulations are performed for the month of September.}
\vspace{-0.3cm}
\label{fig:sim-results-edr-fairness}
\end{figure*}

To capture seasonal variations in atmospheric conditions and their impact on satellite-to-ground quantum communication, we analyze representative days from four different months of the year: March, June, September and December. For each of these dates, we generate location-specific atmospheric profiles using MODTRAN’s AGT. Because clear line-of-sight path is critical for free space quantum communication, cloud cover poses a significant challenge. To account for this, we incorporate real-world weather data from Visual Crossing \cite{vc}, a widely used platform to obtain global historical weather and climate data. Specifically, we use cloud coverage data from the year 2022, aligned with the chosen dates, times of day, and geographic locations, to compute our transmissivity estimates.

We model time of the day effect by treating background photon arrival as equivalent to detector dark clicks. As described in Section \ref{sub:noise}, the presence of background photons decreases the fidelity of the delivered entanglements. We use the following expression to compute background photon flux \cite{harney2022end} at ground station $g$ and approximate it as detector dark click probability.
\begin{align}
    p_g = \frac{H_g\Delta t\Delta\lambda\Omega_g^{\text{fov}}(\pi r_{gR}^2)}{hc/\lambda},
\end{align}
where $H_g$ is the solar irradiance observed by ground station $g$ in units of $\mu$W cm-2 sr-1/nm, and $h$ and $c$ are Planck's constant and speed of light, respectively. $\Omega_g^{\text{fov}}$ and $r_{gR}$ are the field of view and the radius of the receiving telescope at ground station $g.$ In our simulations, we set $\Delta t = 1ns, \Delta\lambda = 1nm, \Omega_g^{\text{fov}} = 10^{-10} sr$ and $r_{gR} = 1m.$ We use MODTRAN to compute $H_g$ for different ground stations at four distinct times of the day $(00.00, 06.00, 12.00 \text{ and } 18.00)$ across four different chosen dates.

We present all of our collected weather data across different ground stations in Figure \ref{weather}. The x-axis lists the ground stations, with the first thirteen located in the northern hemisphere and the remaining four in the southern. In Figure \ref{weather}a, which shows average daily atmospheric transmissivity ($\eta_{g}^{(a)}$), we see a clear shift in values when moving from northern to southern hemisphere ground stations. This is because transmissivity decreases with water vapor concentration. This concentration is generally high in summer and autumn months \cite{zhang2021atmospheric} compared to winter and spring leading to a decrease in transmissivity due to high absorption and scattering. Since seasons are reversed between hemispheres, we see contrasting trends. For example, June has the lowest transmissivity for northern ground stations, but highest for southern ground stations. 

We plot the average cloud coverage across ground stations in Figure \ref{weather}b. The cloud cover values vary widely across stations and seasons, ranging from $0\%$ to $100\%$. We plot the average solar irradiance ($H_g$) in Figure \ref{weather}c. Again, we observe seasonal patterns, values generally peaking in summer and dropping to its lowest in winter.

\begin{figure}[ht]
    \centering
    \begin{subfigure}{0.49\columnwidth}
        \centering
        \includegraphics[width=\linewidth]{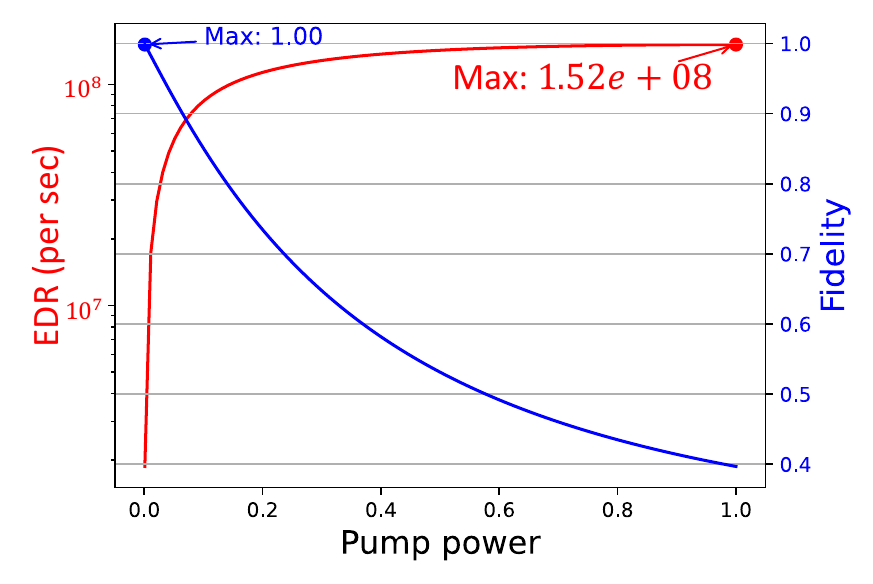} 
        \caption{}
    \end{subfigure}
    \hfill
    \begin{subfigure}{0.49\columnwidth}
        \centering
        \includegraphics[width=\linewidth]{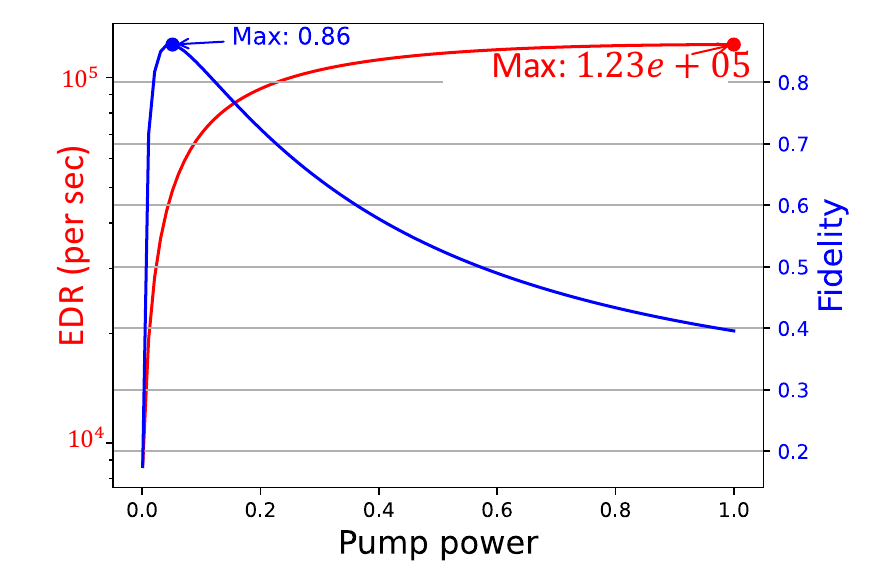}
        \caption{}
    \end{subfigure}
    \caption{Tradeoff between EDR and fidelity. (a) EDR-fidelity tradeoff observed at the entanglement source, reflecting the impact of source parameters (pump power) on rate and fidelity. (b) EDR–fidelity tradeoff observed at the ground station receivers, capturing the effects of source parameters, channel losses, background noise, and detection inefficiencies on the received entangled states.}
    \label{rate-fid-tradeoff}
\end{figure}
\subsection{EDR and Fidelity tradeoff at the source}
Another important factor that influences EDR is the tradeoff between entanglement generation rate and fidelity at the source. The pump power or the mean photon number per mode ($N_s$) can be controlled at the satellite SPDC source affecting the rate and fidelity of the generated entanglements. In Figure \ref{rate-fid-tradeoff}, we show the tradeoff between rate and fidelity as a function of pump power at the source (Figure \ref{rate-fid-tradeoff}a) and once entanglement is achieved between ground stations (Figure \ref{rate-fid-tradeoff}b). An increase in the pump power leads to an increase in EDR at the cost of reduced fidelity due to an increase in the multi-photon events. Thus, depending on the application requirement and tolerance, the pump power should be carefully chosen. In this work, we set pump power to $0.0078$, consistent with the state-of-the-art~\cite{Krovi2016-my} and achieving a fidelity of $0.99$ at the source. However, one can solve a joint optimization problem to dynamically adjust pump powers and determine the optimal assignment strategy to maximize an application specific metric. We view this as a direction for our future work.

\begin{figure}[htbp]
\centering
\begin{minipage}{0.5\textwidth}
\includegraphics[width=1\textwidth]{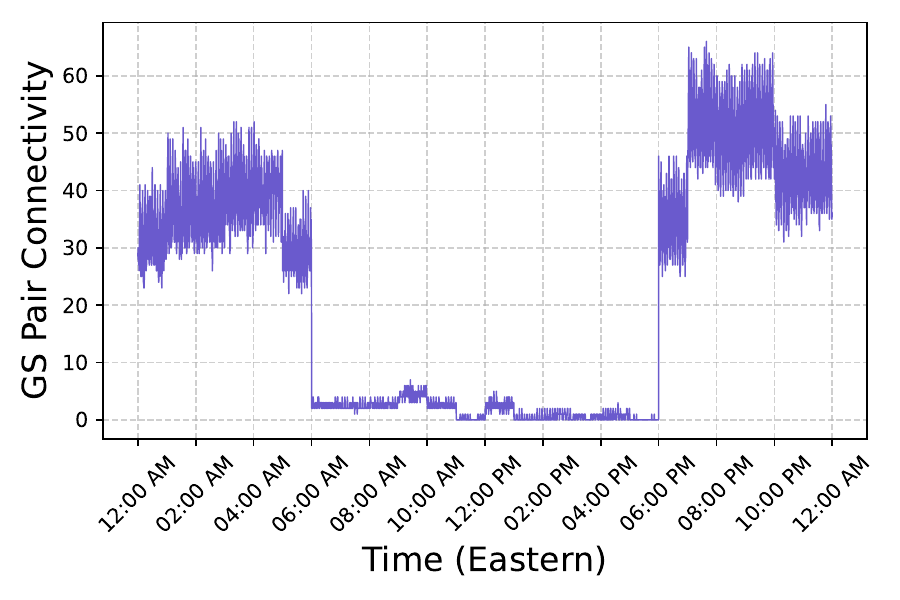}
\end{minipage}\hfill
\caption{Ground station (GS) pair connectivity over time.
Connectivity is defined as the number of GS pairs that, at a given time, have at least one satellite in view capable of delivering entanglement at or above the required fidelity threshold. As satellites orbit the Earth, their visibility to different ground station pairs changes throughout the day. This results in dynamic fluctuations in connectivity, with peaks occurring when many satellites are simultaneously visible, and drops during periods of limited visibility.}
\label{fig:connectivity}
\end{figure}

\begin{figure*}[htbp]
\centering
\begin{minipage}{0.4\textwidth}
\includegraphics[width=1\textwidth]{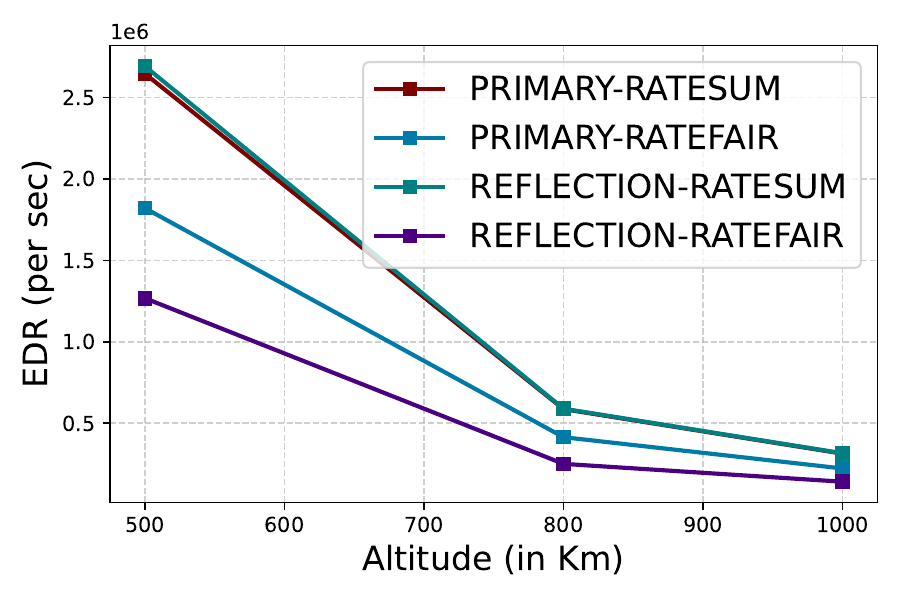}
\subcaption{}
\end{minipage}
\begin{minipage}{0.4\textwidth}
\includegraphics[width=1\textwidth]{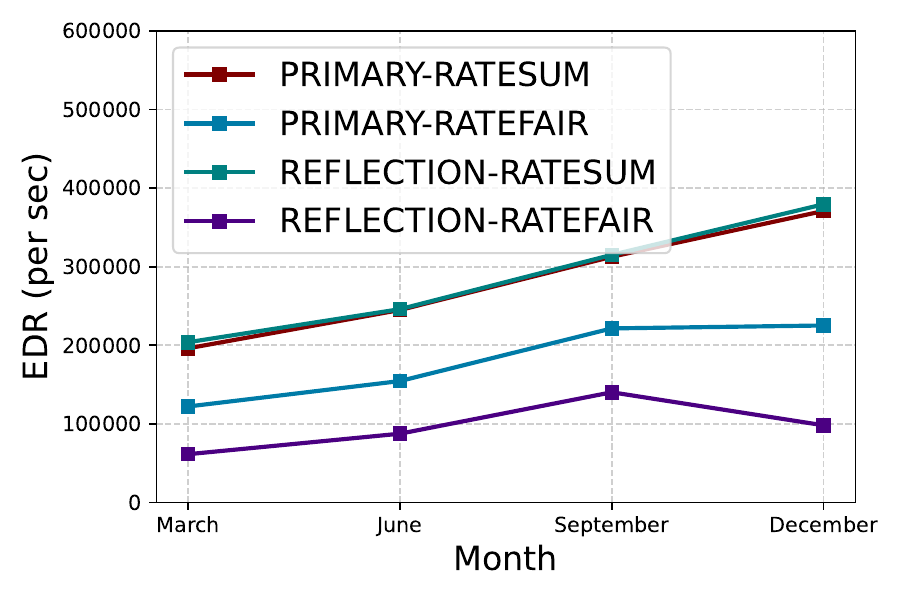}
\subcaption{}
\end{minipage}
\begin{minipage}{0.4\textwidth}
\includegraphics[width=1\textwidth]{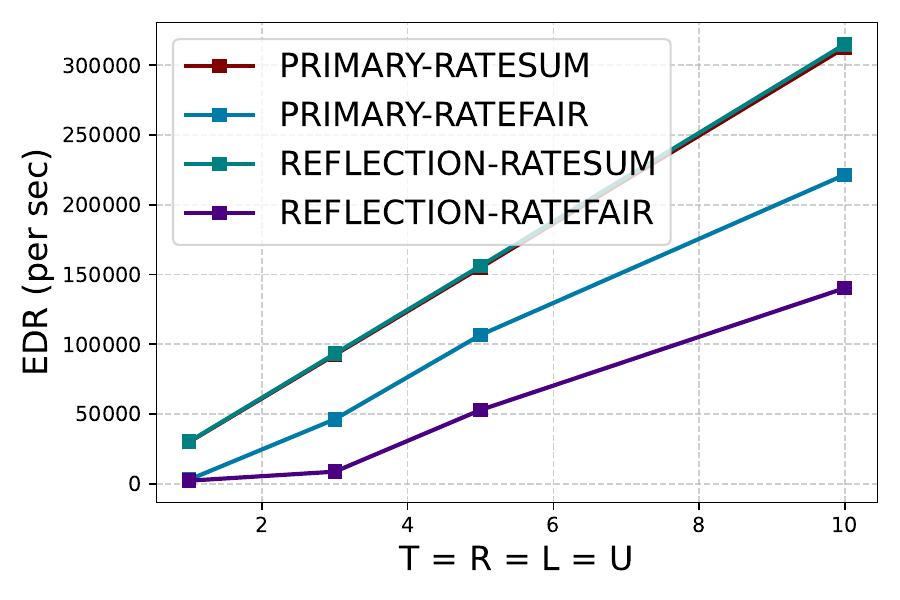}
\subcaption{}
\end{minipage}
\begin{minipage}{0.4\textwidth}
\includegraphics[width=1\textwidth]{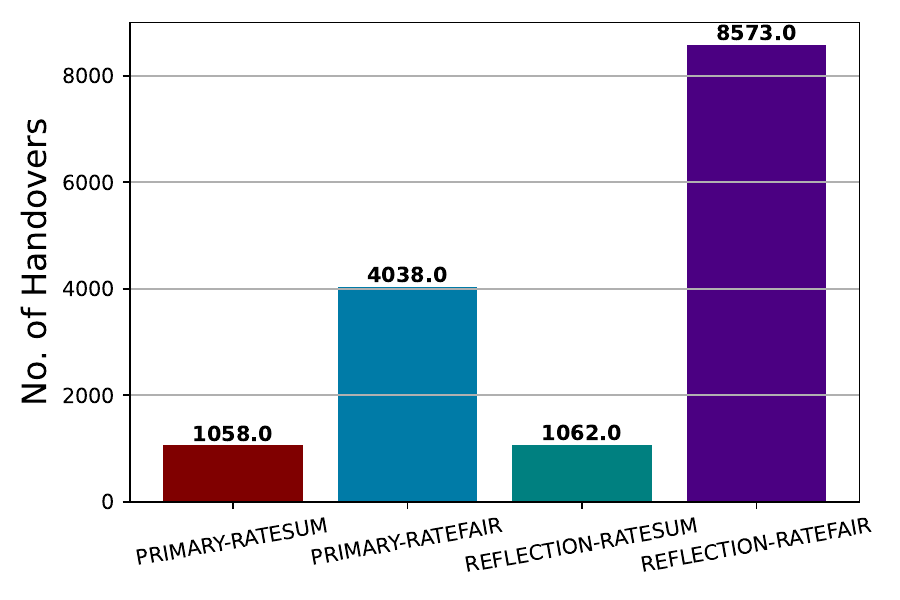}
\subcaption{}
\end{minipage}\hfill
\caption{Performance comparison of four scheduling policies. (a) Impact of satellite orbit altitude on aggregate EDR. (b) Impact of weather conditions on aggregate EDR. (c) Relationship between the number of available resources and EDR (where $T, R,$ and $L$ represent satellite-to-pair, ground station-to-satellite, and pair-to-satellite allocation limits, respectively). (d) Number of satellite handovers observed under each scheduling policy. For plots (b), (c) and (d), the satellite orbit altitude is fixed at $1000$ Km, $T = R = L = 10$ and the simulations are conducted for the month of September.}
\label{fig:sim-others}
\end{figure*}

\subsection{EDR over Time}
We plot the temporal evolution of EDR across four scheduling policies in Figure \ref{fig:sim-results-edr-fairness}(a). The plots reveal oscillations and multi-phase behavior throughout the 24-hour cycle across all scheduling policies. These patterns are mainly due to how GS pair connectivity changes over time as shown in Figure \ref{fig:connectivity}. In Figure \ref{fig:connectivity}, we plot the GS pair connectivity\footnote{We define GS pair connectivity at any given time as the number of ground station pairs that have at least one satellite in view capable of delivering entanglements at the required minimum fidelity.} of the network as time progresses. As satellites move along their orbits, they become visible/hidden to different ground station pairs at different points in the day. This leads to periods where many satellites are visible, resulting in a spike in connectivity, followed by quieter periods when fewer GS pairs are connectable. 

In addition to the oscillations in connectivity, we also observe $5-6$ distinct phases\footnote{A phase refers to a time interval when the moving average of GS pair connectivity remains approximately the same.}. During the 12-hour period from 6:00 AM to 6:00 PM Eastern Time, the network exhibits its lowest GS pair connectivity. This drop is primarily due to high solar irradiance at most ground stations during daylight hours. Elevated background photons significantly degrade the fidelity of the delivered entanglement pairs from satellites to ground stations. As a result, even if satellites are visible, the resulting entanglement fidelity falls below the acceptable threshold ($F^{\text{th}} = 0.85$). These connections are therefore excluded from connectivity calculations, leading to a reduction in effective GS pair connectivity during this time window.

From Figure \ref{fig:sim-results-edr-fairness}(a), we also observe that both \prs\;and \rrs\;achieve high total EDR values, with \rrs\;performing slightly better. This observation aligns with our earlier single ground-station pair case study discussed in Section \ref{sec:reflection}, where \rrs\;demonstrated clear benefits as the baseline distance increased. In the single GS pair setting with only two satellites, the primary scheme often failed to generate entanglements due to lack of simultaneous satellite visibility at both stations, while the reflection-based scheme could still generate entanglement. However, in the current multi-GS pair with satellite constellation scenario, this advantage becomes less significant. Here, \prs\;can leverage diversity and {\it average out} over multiple pairs: even if a few pairs have poor/no visibility at a given time, others will have better visibility, allowing the scheduler to maintain a high overall EDR. In contrast, the fairness based policies (specifically \rrf) prioritizes  scheduling less favorable GS pairs to ensure fairness, which leads to a drop in total EDR.
\subsection{Fairness}
We plot the EDR achieved by each ground station pair across different scheduling policies in Figure \ref{fig:sim-results-edr-fairness}(b). The fairness characteristics of the four scheduling policies can be inferred from the number of ground-station pairs that receive non-zero EDR. \prs, which maximizes total EDR, performs the worst in terms of coverage, serving only $21$ pairs. \prf\;improves upon this, increasing the count to $23$. 
\rrs\;serves $31$ pairs, surpassing \prf\;not just in total EDR but also in fairness. As expected, \rrf\;achieves the highest fairness overall, serving $51$ ground-station pairs with relatively balanced individual EDR values. This is consistent with its design objective of maximizing the minimum fractional EDR across all pairs. 
\subsection{Effect of Orbit Altitude}
We also studied the effect of satellite orbit altitude on average EDR as shown in Figure \ref{fig:sim-others}(a). The overall trends remain consistent, i.e. \rrs\;achieves the best average EDR, closely followed by \prs\;, while \rrf\;provides the best fairness across ground-station pairs (fairness results not shown). At 500 km altitude, all policies perform better due to shorter link distances and lower transmission loss. As altitude increases to 800 km and 1000 km, the average EDR decreases for all four policies. This suggests that lower-altitude constellations are better for entanglement distribution, although they may offer smaller coverage and shorter visibility windows.
\subsection{Seasonal Effects}
The performance of scheduling policies also varies across seasons (see Figure \ref{fig:sim-others}(b)). Among all months considered, December shows the highest EDR for RATESUM policies and \prf, likely because most ground stations are located in the northern hemisphere, where nights are longer during winter. Extended nighttime periods reduce background photons, thereby improving entanglement distribution. This observation is further supported by the atmospheric transmissivity and solar irradiance patterns shown in Figures \ref{weather}(a) and \ref{weather}(c), respectively. Both metrics indicate that December provides the most favorable optical conditions for entanglement distribution: low atmospheric attenuation and minimal background photon flux, making it the best month for entanglement distribution across the majority of ground stations. Despite these seasonal fluctuations in average EDR, all four policies preserve their relative rankings across seasons.
\subsection{Effect of Number of Resources}
We show the effect of increasing number of resources (such as the number of onboard transmitters/sources or receivers) on the average EDR in Figure \ref{fig:sim-others}(c). Increasing the number of resources improves performance for all policies. RATESUM policies benefit more from the increase, while fairness based policies use the added capacity to provide service to the weakest GS pairs, thereby improving overall fairness.

\subsection{Handovers}
One downside of the fairness-based scheduling policies is their higher operational complexity, largely due to the increased number of handovers. We define a handover as an event where an ongoing connection between a satellite and a ground-station pair is transferred to a different satellite due to scheduling decisions. As shown in Figure \ref{fig:sim-others}(d), fairness based policies, in particular \rrf, cause a much larger number of satellite handovers compared to RATESUM policies. This is because fairness-based scheduling frequently reallocates satellite resources to serve less favorable ground-station pairs, which often requires switching connection among multiple satellites as visibility conditions change. RATESUM policies prioritize high EDR connections. They maintain connection with the most favorable ground-station pairs for longer durations, resulting in fewer handovers. Also, note that, \rrs\;performs nearly as well as \prs\;in terms of the number of handovers.

%% file: conclusion.tex
\section{Conclusion and Open Problems}\label{sec:conclusion}
In this work, we presented an optimization based framework for scheduling in quantum satellite networks that jointly considers resource constraints, environmental factors and fairness. By formulating the scheduling problem as an integer linear program, we evaluated multiple objectives: maximizing entanglement distribution rate and ensuring fair access across ground station pairs. Our model incorporates realistic entanglement sources based on SPDC and captures channel impairments due to atmospheric conditions and background noise. Furthermore, we addressed the added complexity of coordinated entanglement distribution through two-satellite relays, proposing an optimal scheduling strategy for this scenario. Going further, we aim to extend our analysis to characterize optimal scheduling policy for multi-partite entanglement distribution. A simulation study to compare and contrast the performance of different satellite constellations remain one of our future works.